%% file: main.tex
\documentclass[conference]{IEEEtran}

\usepackage[utf8]{inputenc}
\usepackage[T1]{fontenc}
\usepackage{textcomp}     
\usepackage{fancyhdr}
\usepackage{times}
\usepackage{listings}
\usepackage{amssymb}

\usepackage{amsmath}
\usepackage{array}
\usepackage{url}
\usepackage{pgfplots}

\usepackage{epstopdf}
\usepackage{tikz}
\usetikzlibrary{snakes}
\usetikzlibrary{shapes,arrows}
\usepackage{multirow}
\usepackage{booktabs}
\usepackage{graphicx}

\usepackage{subcaption}

\usepackage{verbatim}
\usepackage[font=small,labelfont= bf]{caption}
\usepackage{ifsym}
\usepackage[algo2e,vlined,linesnumbered,ruled,noend]{algorithm2e}  
\SetKwRepeat{Do}{do}{while}%
 
\usepackage{relsize}
\usetikzlibrary{shapes,arrows}
\usetikzlibrary{patterns}
\usepackage{color,soul}
\newcommand{\removelatexerror}{\let\@latex@error\@gobble}

\usepackage{pifont}
\newcommand{\cmark}{\ding{51}}%

\usetikzlibrary{shapes,positioning,fit,calc}
\usetikzlibrary{trees}
\usetikzlibrary{arrows}
\tikzstyle{snakeline} = [decorate, decoration={snake, amplitude=.4mm,
                         segment length=2mm},thick]
\tikzset{
  treenode/.style = {align=center, inner sep=0pt, text centered,font=\sffamily},
  arn_n/.style = {treenode, circle,black, draw=black, text width=1.3em}, 
  rrNode/.style = {draw, circle,node distance=0.8cm}
}

\def\bsq#1{\textquotesingle #1\textquotesingle}

\DeclareCaptionType{copyrightbox}

\newcommand{\kat}[1]{ {\mbox{\tiny #1}} }

\newcommand{\LJoin}{\tiny\textifsym{d|><|}}
\newcommand{\RJoin}{\tiny\textifsym{|><|d}}
\newcommand{\OJoin}{\tiny\textifsym{d|><|d}}
\newcommand{\AJoin}{\tiny\rhd}

\newcolumntype{M}[1]{>{\centering\arraybackslash}m{#1}}
\newcolumntype{L}[1]{>{\raggedright\let\newline\\\arraybackslash\hspace{0pt}}m{#1}}

\newcommand\VRule[1][\arrayrulewidth]{\vrule width #1}
\newcolumntype{C}[1]{>{\centering\let\newline\\\arraybackslash\hspace{0pt}}m{#1}}

\definecolor{c1}{RGB}{75,115,75}
\definecolor{c2}{RGB}{60,116,131}
\definecolor{c3}{RGB}{157,50,50}
\definecolor{c4}{RGB}{0,177,179}
\definecolor{c5}{RGB}{212,175,55}
\definecolor{grey}{RGB}{169,169,169}

\newtheorem{definition}{Definition}

\newtheorem{example}{Example}

 \newcommand{\squishlist}{
 \begin{list}{$\bullet$}
  { \setlength{\itemsep}{0pt}
     \setlength{\parsep}{1pt}
     \setlength{\topsep}{1pt}
     \setlength{\partopsep}{0pt}
     \setlength{\leftmargin}{1.5em}
     \setlength{\labelwidth}{1em}
     \setlength{\labelsep}{0.5em} } }
 \newcommand{\squishend}{\end{list}}
 
\usepackage{todonotes}

\def\HiLi{\leavevmode\rlap{\hbox to \hsize{\color{yellow!50}\leaders\hrule height .8\baselineskip depth .5ex\hfill}}}

\usepackage[inline]{enumitem}

\def\Hili{\leavevmode\rlap{\hbox to
    \hsize{\color{yellow!50}\leaders\hrule height .8\baselineskip
      depth .5ex\hfill}}}

\begin{document}
        
\title{Generalized Lineage-Aware Temporal Windows: Supporting Outer and Anti Joins\\ in Temporal-Probabilistic Databases}

\author{\IEEEauthorblockN{Katerina Papaioannou\IEEEauthorrefmark{1},
Martin Theobald\IEEEauthorrefmark{2},
Michael B{\"o}hlen\IEEEauthorrefmark{1}}
\vspace*{0.15cm}
\IEEEauthorblockA{\IEEEauthorrefmark{1} \em Department of Computer Science --
University of Zurich\\ $\mathtt{\{papaioannou,boehlen\}@ifi.uzh.ch}$}
\vspace*{0.15cm}
\IEEEauthorblockA{\IEEEauthorrefmark{2} \em Faculty of Science, Technology \& Communication -- University of Luxembourg \\ $\mathtt{martin.theobald@uni.lu}$ }}


\maketitle

\begin{abstract}
The result of a temporal-probabilistic (TP) join with negation
includes, at each time point, the probability with which a
tuple of a positive relation ${\bf p}$ matches {\em none} of the tuples in
a negative relation ${\bf n}$, for a given join condition $\theta$.
%
%
TP outer and anti joins thus resemble the characteristics of relational outer and anti joins also in the case when there exist time points 
at which input tuples from ${\bf p}$ have non-zero probabilities to be $true$ and input tuples from ${\bf n}$ have non-zero probabilities to be $false$, respectively.
%
%
For the computation of TP joins with negation, we introduce
{\em generalized lineage-aware temporal windows}, a mechanism that
binds an output interval to the lineages of all the matching valid
tuples of each input relation.
We group the windows of two TP relations into three disjoint
sets based on the way attributes, lineage expressions
and intervals are produced.
We compute all windows in an incremental manner, and we show
that pipelined computations allow for the direct integration of our
approach into PostgreSQL. We thereby alleviate the prevalent redundancies in the interval computations
of existing approaches, which is proven by an extensive experimental
evaluation with real-world datasets.
%
%
%
%
\end{abstract}

\section{Introduction}
\input{1_introduction}

\section{Related Work}
\label{sec:related-work}
\input{2_related_work}

\section{Background}
\label{sec:background}
\input{3_background}

\section{Negation in TPDBs}
\label{sec:properties}
\input{4_operation_properties}

\section{Generalized Windows}
\label{sec:windows-defined}
\input{5_windows}

\section{Algorithms}
\label{sec:algorithms}
\input{7_algorithms}

\section{Evaluation}
\label{sec:exper-eval}
\input{8_experimental_evaluation}
\section{Conclusions}
\label{sec:conclusion}

In this work, we proposed an approach for the computation of
temporal-probabilistic joins with negation, operations that
cannot currently be performed by any existing TP approach. 
We introduced the generalized lineage-aware temporal windows,
to bind lineages and intervals and comply with the requirements
of TP joins. We grouped these windows into three sets and, using
these sets, we expressed the result of each TP join with negation.
We implemented algorithms for the pipelined computation of all
sets of generalized lineage-aware temporal windows and we
integrated our approach in the kernel of PostgreSQL. A thorough
experimental evaluation reveals that our implementation is
seamlessly integrated into the DBMS and outperforms existing
approaches.

\bibliographystyle{IEEEtran}
\bibliography{IEEEabrv,citation}


\end{document}

%% file: 1_introduction.tex
Join operations with negation are performed for a positive
relation ${\bf p}$, a negative relation relation ${\bf n}$ and a
$\theta$ condition that determines the tuples that match. In
conventional databases, joins with negation disqualify an
input tuple of the positive relation if its attributes match the
attributes in a tuple of the negative relation. In temporal
databases, the existence of a matching tuple in ${\bf n}$ does
not disqualify the tuple of ${\bf p}$ itself but timepoints at
which it is valid~\cite{BohlenBJ98, BohJen2009}. In probabilistic
databases, where tuples have a probability to be true or false,
the existence of a matching tuple in ${\bf n}$ only reduces the
probability with which a tuple is included in the
output~\cite{Suciu2009, Wang2008}. 

The result of a temporal-probabilistic join with negation
includes, at each time point, the probability with which a tuple
of the positive relation ${\bf p}$ matches no tuple in the
negative relation ${\bf n}$ for a predicate $\theta$. 
Firstly, it includes output tuples that span subintervals when
only tuples of ${\bf p}$ are valid. In such cases, output
intervals might be determined by starting or ending points of
input tuples that are not valid during the output interval.
Secondly, TP joins with negation produce outputs
that indicate, at each time point, the probability of a tuple
$\tilde{p}$ in ${\bf p}$ not matching any valid tuple
in ${\bf n}$ because all of them are false. In this case, an
output interval $T$ is determined based on the starting and
ending points of $\tilde{p}$ and of the tuples of ${\bf n}$
that are valid over $T$ and match $\tilde{p}$ for $\theta$.


\begin{figure}[!h]
\fontsize{8pt}{11pt}\selectfont
\setlength{\tabcolsep}{3pt}
\centering 
\begin{subfigure}[b]{\linewidth}
\centering
\scalebox{1}{
\begin{tabular}{ c c | c | c | c}
\multicolumn{5}{l}{\bf a (wantsToVisit)}\\
\hline
{\em Name} & {\em Loc} & $\lambda$ & $T$      & $p$\\ 
\hline
Ann        & {ZAK}    		& ${a_1}$   & {[2,8)}  & 0.7\\    
Jim        & {WEN}       	& ${a_2}$   & {[7,10)} & 0.8\\     
\hline
\end{tabular}
}      
\qquad
\scalebox{1}{
\begin{tabular}{ c c | c | c| c }
\multicolumn{5}{l}{\bf b (hotelAvailability)}\\
\hline
{\em Hotel} & {\em Loc} & $\lambda$  &  $T$    & $p$\\ \hline
{hotel$_3$} & {SOR}  	& ${b_1}$    & {[1,4)} & 0.9\\ 
{hotel$_2$} & {ZAK} 	& ${b_2}$    & {[5,8)} & 0.6\\ 
{hotel$_1$} & {ZAK} 	& ${b_3}$    & {[4,6)} & 0.7\\
\hline
\end{tabular}}
\caption{Temporal-probabilistic base relations}
\label{fig:relationsTPDB}
\end{subfigure}
\vspace{0.1cm}

\begin{subfigure}[b]{\linewidth}
\centering 
\scalebox{1}{
\begin{tabular}{ c c c | c | c | c }
\multicolumn{6}{l}{Q = $\bf a \ \LJoin^\kat{Tp}_\theta \ \ b$, 
\ $\theta : \mathbf{a}$.Loc = $\mathbf{b}$.Loc}\\
\hline
$Name$ & $Loc$    & $Hotel$    & $\lambda$          & $T$      & $p$\\ 
\hline
Ann    & ZAK   & -          & $a_1$            		& {[2,4)} & 0.70\\ 
Ann    & ZAK   & hotel$_1$  & $a_1 \land b_3$  		& {[4,6)} & 0.49\\ 
Ann    & ZAK   & hotel$_2$  & $a_1 \land b_2$  		& {[5,8)} & 0.42\\
Ann    & ZAK   & -          & $a_1 \land \lnot b_3$ & {[4,5)} & 0.21\\
Ann    & ZAK   & -          & $a_1 \land \lnot (b_3 \lor b_2)$
	   & {[5,6)} & 0.084\\
Ann    & ZAK   & -          & $a_1 \land \lnot b_2$ & {[6,8)} & 0.28\\
Jim    & WEN   & -          & $a_2$			 		& {[7,10)}& 0.80\\
\hline
\end{tabular}}
\caption{Temporal-probabilistic tuple-based query}
\label{fig:exampleQuery}
\end{subfigure}
\caption{Temporal-probabilistic database example}
\label{fig:basicTPDBsuper}
\end{figure}

\begin{example}
Consider a booking website (Figure~\ref{fig:basicTPDBsuper}) that
archives prediction data over time. Table $\mathbf{a}$ records 
data related to the locations that the clients want to visit,
according to their searches. Table $\mathbf{b}$ records data 
regarding the availability of the hotels registered in the 
website, considering the busy periods in each location and
the rate at which each hotel gets booked. This archive
corresponds to a temporal-probabilistic database. Tuple
(\bsq{Jim, WEN}, $a_2$, [7,10), 0.8) captures that, at each day
from the $7^{th}$ to the $10^{th}$ of the month, \bsq{Jim wants
to visit Wengen} with probability {\it 0.8}. The website makes a
prediction for each time point and there is no other tuple in
$\mathbf{a}$  that predicts the probability of \bsq{Jim visiting
Wengen} over an interval overlapping with [7,10). In order to
manage supply and demand, we determine the probability with which
the client will find available accommodation at their preferred 
location, at each time point. The corresponding query is Q =
$\bf a \ \LJoin^\kat{Tp}_\theta \ \ b$ ($\theta : \mathbf{a}$.Loc
= $\mathbf{b}$.Loc), i.e., a temporal-probabilistic outer join
with equality on the locations.

The answer tuple (\bsq{Ann, ZAK, hotel$_1$}, $a_1 \land b_3$,
[4,6), 0.49) expresses that, with probability $0.49$, Ann wants to
visit Zakynthos ($a_1$) and stay at hotel$_1$ in Zakynthos 
($b_3$) during interval [4,6). It is valid over the intersection
of the intervals of tuples $a_1$ and $b_3$ and it is true when
both these tuples are true. Answer tuple (\bsq{Ann, ZAK, -},
$a_1$, [2,4), 0.7) expresses that, with probability $0.7$, Ann
wants to visit Zakynthos ($a_1$) but there is no hotel available
to stay there. Although the lineage and the output probability are
both determined by tuple $a_1$, i.e., the only tuple valid during
[2,4), the interval of this output tuple is influenced by the
starting point of tuple $b_3$, a tuple not valid over [2,4). Over
the interval [5,6) there is $0.084$ probability that Ann wants to 
visit Zakynthos but finds no accommodation. According to answer
tuple (\bsq{Ann, ZAK, - }, $a_1 \land \lnot (b_3 \lor b_2)$,
[5,6), 0.084), during [5,6), the output is influenced by more than
a pair of input tuples. Although all tuples are valid over [5,6),
this tuple is true when \bsq{Ann visits Zurich} ($a_1$ is true)
but also when neither hotel$_1$ nor hotel$_2$ are available during
[5,6) ($b_3$ and $b_2$ are false). 
\end{example}

TP set-difference is the only temporal-probabilistic operation with
negation that has been investigated~\cite{papaioannou2018}.  Since
set-operations combine only tuples with equal non-temporal attributes,
simplified structures can be used. Specifically, only one tuple of
each relation is valid at each time point, which allows for solutions
with linearithmic complexity. For TP outer joins and TP anti join,
multiple tuples of the negative relation might be valid over an output
interval and input tuples with non-temporal attributes that are not
pairwise equal might be combined to form an output tuple. Moreover, TP
outer joins combine the characteristics of TP joins with and without
negation: at each time point, two outcomes are possible since the same
tuples can be $true$ or \textit{false}.

\begin{figure}[!h]
\fontsize{8pt}{11pt}\selectfont
\setlength{\tabcolsep}{4pt}
\centering 
\begin{subfigure}[b]{\linewidth}
\centering 
\scalebox{1}{
\begin{tabular}{ c c | c | c | c | c }
\cline{2-6}
      & $F_r$          & $F_s$ 	& $\lambda_r$ & $\lambda_s$ & $T$     \\ \cline{2-6}
$w_1$ & \bsq{Ann, ZAK} & -      & $a_1$		  & -           & {[2,4)} \\ 
$w_2$ & \bsq{Jim, WEN} & -      & $a_2$		  & -  	        & {[7,10)} \\
\cline{2-6}
\end{tabular}}
\vspace*{0.1cm}
\caption{Unmatched Windows}
\label{fig:introUnmatched}
\end{subfigure}

\vspace*{0.2cm}

\begin{subfigure}[b]{\linewidth}
\centering 
\scalebox{1}{
\begin{tabular}{ c c |  c | c | c | c }
\cline{2-6}
      & $F_r$          & $F_s$ 			    & $\lambda_r$ & $\lambda_s$ & $T$ \\ \cline{2-6}
$w_3$ & \bsq{Ann, ZAK} & \bsq{hotel$_1$, ZAK} & $a_1$		& $b_3$       & {[4,6)} \\ 
$w_4$ & \bsq{Ann, ZAK} & \bsq{hotel$_2$, ZAK} & $a_1$		& $b_2$  	  & {[5,8)} \\
\cline{2-6}
\end{tabular}}
\vspace*{0.1cm}
\caption{Overlapping Windows}
\label{fig:introOverlapping}
\end{subfigure}

\vspace*{0.2cm}

\begin{subfigure}[b]{\linewidth}
\centering 
\scalebox{1}{
\begin{tabular}{ c c c |c |c | c  }
\cline{2-6}
      & $F_r$          & $F_s$ 	& $\lambda_r$ & $\lambda_s$    & $T$     \\ \cline{2-6}
$w_5$ & \bsq{Ann, ZAK} & -      & $a_1$ 	  & $b_3$          & {[4,5)} \\
$w_6$ & \bsq{Ann, ZAK} & -      & $a_1$ 	  & $b_3 \lor b_2$ & {[5,6)} \\
$w_7$ & \bsq{Ann, ZAK} & -      & $a_1$ 	  & $b_2$ 		   & {[6,8)} \\
\cline{2-6}
\end{tabular}}
\vspace*{0.1cm}
\caption{Negating Windows}
\label{fig:introNegating}
\end{subfigure}
\caption{Generalized lineage-aware temporal windows of relations ${\bf a}$ and ${\bf b}$ (Fig.~\ref{fig:relationsTPDB}) for the
$\theta$-condition a.Loc=b.Loc}
\label{fig:basicTPDBsets}
\end{figure}


\smallskip
\noindent\textbf{Outline \& Contributions.}
\squishlist

\item We introduce {\em generalized lineage-aware temporal windows} to
  produce output tuples for input pairs with different non-temporal
  attributes and for cases when multiple input tuples are valid. Given
  a $\theta$-condition and two TP relations, we group windows into
  three disjoint sets: the {\em unmatched}, the {\em overlapping} and
  the {\em negating windows}.  An output tuple is formed for each
  window using the appropriate lineage-concatenation functions and we
  express the result of TP joins with negation using the three sets.

\smallskip
\item We introduce the algorithms LAWA$_\kat{U}$ and LAWA$_\kat{N}$
  for the computation of unmatched and negating windows,
  respectively. Recording the lineages of the tuples valid in each
  input relation over an output interval and keeping them decoupled
  until the formation of output tuples, allows for the computation of
  unmatched and negating windows based on the overlapping ones. Thus,
  redundant interval comparisons due to the repetition of basic steps
  are avoided and the runtime required for the computation of outer
  joins and anti join improves by two orders of magnitude.

\smallskip
\item We conduct extensive experiments using real datasets to compare
  our approach for the computation of TP outer joins and TP anti join
  with existing state of the art approaches. Our approach is
  integrated in PostgreSQL and exhibits a lower runtime while being
  scalable.  \squishend

  \smallskip The remainder of this paper is organized as follows.
  Section~\ref{sec:related-work} provides an overview of related works
  on temporal and probabilistic databases with a focus on outer joins
  and anti join. Section~\ref{sec:background} discusses the TP data
  model and its query semantics.  Section~\ref{sec:properties}
  discusses the impact of negation in TP
  joins. Section~\ref{sec:windows-defined} introduces generalized
  lineage-aware temporal windows and groups them into three disjoint
  sets. Section~\ref{sec:algorithms} introduces two algorithms for the
  computation of the different window sets while
  section~\ref{sec:exper-eval} presents a comprehensive performance
  study that compares our implementation with existing approaches.
  Section~\ref{sec:conclusion} concludes the paper.

%% file: 2_related_work.tex

We review related approaches from temporal and probabilistic
databases and explain their limitations in terms of supporting TP
outer joins and anti join.

\smallskip \noindent\textbf{Temporal-Probabilistic Operations.}
Dylla et al.\ \cite{DyllaMT13} introduced a closed and complete
TP database model, coined TPDB, based on existing temporal and
probabilistic models. Query processing is performed in two
steps. The first step, grounding, evaluates a chosen deduction
rule (formulated in Datalog with additional time variables and
temporal predicates) and computes the lineage expressions of the
deduced tuples. The second step, deduplication, removes the
duplicates that could occur in the grounding step by adjusting
intervals. The grounding step performs pairwise
tuple-comparisons. Subintervals that are present in only one of
the two input relations, i.e., during which no tuple of the other
relation is valid, cannot be produced.

\smallskip \noindent\textbf{TP Operations with negation.}
Set-difference is the only TP operation with negation that has been
investigated~\cite{papaioannou2018}. For its computation, Papaioannou
et al. introduced {\em lineage-aware temporal windows}, a mechanism
that binds an output interval with the lineage of the tuple in each
input relation that includes fact $F$ and that is valid during the
interval.  {\em Lineage-aware temporal windows} eliminate redundant
interval comparisons and additional joins for the formation of lineage
expressions in TP set operations. The starting and ending points of
the interval that the window spans are computed via a comparison of
the starting and ending points of input tuples that are valid but also
of neighboring tuples. Thus, they are useful for output intervals that
are not equal to the overlap of a pair of valid tuples. However, they
are tailored to cases when one tuple of each input relation is valid
and when the input tuples have the same non-temporal attributes. In TP
joins with negation, input tuples with different non-temporal
attributes are combined and multiple tuples of an input relation can
be valid over an interval and need to be included in the lineage of an
output tuple.

\smallskip \noindent\textbf{Temporal Joins.}  In temporal
databases, the result of a temporal outer join $op^T$ is defined
as the result of applying $op$ over a sequence of atemporal
instances (the so-called snapshots) of the input relations---a
key concept in temporal databases termed {\em snapshot
reducibility} ~\cite{AlKatebGCBCP13, lorentzos1997sql, 
Viqueira2007}.  Maximal intervals are produced by merging
consecutive time points to which the same input tuples have
contributed ({\em change preservation}). Dign\"os et
al.~\cite{DignosBG12, DignosTODS16} use {\em data lineage} to
guarantee change preservation for all relational operations under
a sequenced semantics. For the computation of joins, they
introduce the {\em alignment} operator. The 
alignment $\Phi({\bf r}, {\bf s})$ of a relation ${\bf r}$ based
on another relation ${\bf s}$ replicates the tuples of ${\bf r}$
and assigns new time intervals to them. The new intervals are
obtained by splitting the original intervals of ${\bf r}$ based
on tuples of ${\bf s}$ with which they overlap. The valid tuples
of both relations that contribute to an adjusted interval are not 
recorded. This is the reason why the alignment of both relations
is required as well as the application of $op$ to produce all
output tuples~\cite{DignosBG12,DignosTODS16}. Using this approach
in a TP context, other than the overhead and redundancy of
aligning both relations, the input tuples must also be adjusted in
groups and not only in pairs for the cases when valid tuples are
$false$. Combining adjustment both in pairs and in groups multiple
times in the same query incurs redundant comparisons and 
recomputation of intermediate results.

{\em Sweeping-based approaches} have been widely used for the
computation of overlap joins~\cite{PlatovICDE16, Arge1998} in
temporal settings. A sweepline moves over all start and end 
points of tuples, and determines, for each time point, the tuples
of both input relations that are valid. These approaches are
tailored to compute efficiently the overlap join but are not 
suitable for the computation of the class of operations discussed 
in this paper. First, the overlapping intervals computed in these
approaches only correspond to a part of the result of a TP
outer join while they are not included in the result of a TP
anti join. Second, they generally do not consider join conditions
on the non-temporal attributes limiting the types of queries they
could be used for.


\smallskip\noindent\textbf{Probabilistic Joins.} In
probabilistic databases, the result of a probabilistic operation
$op^p$ is defined as the result of applying $op$ over the set of
all possible instances of the input relations. The Trio
system~\cite{SarmaTW08} was among the first to recognize {\em 
data lineage}, in the form of a Boolean formula, as a means to
capture the possible instances at which an output tuple is valid.
In an effort to provide a {\em closed and complete} 
representation model for uncertain relational data, they 
introduced {\em Uncertainty and Lineage Databases}
(ULDBs)~\cite{BenjellounSHTW08}. The algebraic operators are
modified to compute the lineage of the result tuples in a ULDB,
thus capturing all information needed for computing query answers
and their probabilities.  Fink et al.~\cite{FinkOR11,FinkO16}
reduced the computation of probabilistic algebraic operations to
conventional operations so that these can be performed using a
DBMS, rather than by an application layer built on top of it. In
all these works, the focus is restricted to select-project join
queries. Probabilistic anti join, expressed with the NOT EXISTS 
predicate in SQL, has been explored by Wang et
al.~\cite{Wang2008}. It has been integrated in MystiQ by
breaking the initial query into positive and negative subqueries
that are separately evaluated and then combined. Incorporating
interval computation with predicates in these approaches is
possible but does not comply with all the requirements of TP
operations with negation.



%% file: 3_background.tex

We denote a {\bf temporal-probabilistic schema} by $R^\kat{Tp}$
($F$, $\lambda$, $T$, $p$), where $F$ = ($A_1$, $A_2$, $\ldots$,
$A_m$) is an ordered set of attributes, and each attribute $A_i$
is assigned to a fixed domain $\Omega_i$. $\lambda$ is a Boolean
formula corresponding to a lineage expression. $T$ is a {\em 
temporal attribute} with domain $\Omega^T \times \Omega^T$,
where $\Omega^T$ is a finite and ordered set of {\em time
points}. $p$ is a {\em probabilistic attribute} with domain
$\Omega^p = (0,1] \subset {\rm I\!R}$. A {\bf
temporal-probabilistic relation $\mathbf{r}$} over $R^\kat{Tp}$
is a finite set of tuples. Each tuple $r \in \mathbf{r}$ is an
ordered set of values from the appropriate domains. The value of
attribute $A_i$ of $r$ is denoted by $r.A_i$. The conventional
attributes $F$ = ($A_1$, $A_2$, $\ldots$, $A_m$) of tuple $r$
form a {\em fact}, and we write $r.F$ to denote the fact $f$
captured by tuple $r$. For example, base tuple (\bsq{Ann,
ZAK}, $a_1$, $[2,8)$, $0.7$) of relation $\mathbf{a}$ (see
Fig.~\ref{fig:relationsTPDB}) includes the fact $a_1.F$ =
(\bsq{Ann, ZAK}), the lineage expression $a_1.\lambda = a_1$,
the time interval $a_1.T = [2,8)$, and the probability value
$a_1.p = 0.7$. The temporal-probabilistic annotations of the
schema express that
\begin{enumerate*}
\item[(i)] $a_1 = \mathit{true}$ with probability $a_1.p$ for every
  time point in $a_1.T$,
\item[(ii)] $a_1 = \mathit{false}$ with probability $1-a_1.p$ for
  every time point in $a_1.T$, \item[(iii)] and $a_1$ is always
  $\mathit{false}$ outside $a_1.T$.
\end{enumerate*} 
By following conventions from 
\cite{DyllaMT13,DignosTODS16,DignosBG12,OlteanuHK09}, we assume 
duplicate-free input and output relations. Formally, a
temporal-probabilistic relation $\mathbf{r}$ is {\bf
duplicate-free} iff $\forall r, r' \in \mathbf{r} (r \neq
r' \Rightarrow r.F \neq r'.F \vee r.T \cap r'.T = \emptyset))$. 
In other words, the intervals of any two tuples of $\mathbf{r}$
with the same fact $f$ do not overlap.

A {\bf lineage expression} $\lambda$ is a Boolean formula,
consisting of tuple identifiers and the three Boolean
connectives $\neg$ (``not"), $\land$ (``and") and $\lor$
(``or").  Tuple identifiers represent Boolean random variables
among which we assume independence \cite{DyllaMT13, OlteanuHK09,
DalviS07}. For a base tuple $r$, $r.\lambda$ is an atomic
expression consisting of just $r$ itself. For a result tuple
$\tilde{r}$ derived from one or more TP operations,
$\tilde{r}.\lambda$ is a Boolean expression as defined above.
The probability of a result tuple is computed via a probabilistic
valuation of the tuple's lineage expression, using either exact
(see, e.g., \cite{DalviS07, DalviS12, olteanu2008using}) or
approximate (see, e.g., \cite{FinkHO13,FinkO11,
gatterbauer2014oblivious, GatterbauerS15, OlteanuHK10})
algorithms.  For example, in the result relation of
Fig.~\ref{fig:exampleQuery}, the lineage $a_1 \land \lnot b_3$
yields a marginal probability of $0.7 \cdot (1-0.7) = 0.21$ by
assuming independence among the base tuples $a_1$ and $b_3$ (see
Fig.~\ref{fig:relationsTPDB}).

We write $\lambda^{\mathbf{r}, f}_{t}$ to refer to the
disjunction of the lineage expressions of the tuples in
relation $\mathbf{r}$ with fact $f$ that are valid at time point
$t$. We write $\lambda^{\mathbf{r}, \theta}_{t}$ to refer to the
disjunction of the lineage expressions of the tuples in
relation $\mathbf{r}$ that satisfy $\theta$ and are valid at
time point $t$. When there are no tuples in $\mathbf{r}$ with
fact $f$ or satisfying $\theta$ at time point $t$, we write
$\lambda^{\mathbf{r}, f}_{t} = \mathtt{null}$ or 
$\lambda^{\mathbf{r}, \theta}_{t} = \mathtt{null}$,
respectively. We write $\theta_{\tilde{r}}$ to indicate that
values of attributes in condition $\theta$ are instantiated
to the corresponding values in tuple $\tilde{r}$. For
example, for the $\theta$ condition used in the query of
Figure~\ref{fig:exampleQuery} and $\tilde{r}$ = (\bsq{Ann, ZAK,
hotel$_1$}, $a_1 \land b_3$, [4, 6), 0.49), we get
$\theta_{\tilde{r}}: b.Loc$ = \bsq{ZAK}.



The semantics of the TP data model are centered around two 
properties: TP snapshot reducibility and TP change
preservation~\cite{papaioannou2018}. {TP Snapshot reducibility}
states that the result of $op^\kat{Tp}$ at each time point $t$
is equal to the result of $op^\kat{p}$ on the input tuples with
non-zero probability to be valid at $t$. Thus, the output 
attributes are determined only by the input tuples at $t$ and 
the output lineages and probabilities are consistent with the
possible-worlds semantics~\cite{SarmaTW08, BenjellounSHTW08}.
The TP left outer join of Fig.~\ref{fig:exampleQuery} complies
with TP snapshot-reducibility. For example, in tuple (\bsq{Ann,
ZAK, hotel$_1$}, [4,6), $a_1 \land b_3$, 0.42), at time point
$t=4$, the fact is a combination of $a_1.F$ = \bsq{Ann, ZAK} and
$b_3.F$ = \bsq{hotel$_1$, ZAK}, i.e., the only input tuples valid
at $t$ and whose facts satisfy the join condition. 

TP change preservation ensures that only consecutive time points
of output tuples with equal facts and equivalent lineage 
expressions are grouped into intervals. It guarantees maximal 
intervals where the lineage expression is the same at all time 
points in the interval and different at time points outside. For
example, the output tuples (\bsq{Ann, ZAK, -}, [2,4), $a_1$, 0.7)
and (\bsq{Ann, ZAK, -}, [4,5), $a_1 \land \neg b_3$, 0.42) were
not merged into the interval $[2,5)$, since they do not have equivalent lineages.

%% file: 4_operation_properties.tex
The characterization of joins as operations with and without
negation has been well established in databases \cite{FinkO16}.
As illustrated in Table~\ref{tab:criteria_negation}, the
Cartesian product and the inner join are joins without negation
since they only record information valid in both input relations.
The anti join is a join purely based on negation and
outer joins combine joins with and without negation. 

\begin{table}[!htpb]
\caption{Join Operations Categorized Based on Negation}
\label{tab:criteria_negation}
\centering
\scalebox{0.9}{
\begin{tabular}{ |  M{1.65cm}  !{\VRule[1.5pt]} M{2.5cm} | @{}m{0cm}@{}}
\hline
\specialrule{1.2pt}{0pt}{0pt}
        & \textbf{Operations} & \\ [0.1cm] 
\hline
WITHOUT & $\times$,\ $\bowtie$ & \\ [0.1cm] \hline 
WITH    & $\AJoin$ & \\ [0.1cm] \hline 
MIXED   & $\LJoin$,\ $\RJoin$,\
          $\OJoin$ & \\ [0.1cm] \hline 
\end{tabular}}
\end{table}

%

A join with negation is performed over a positive relation
${\bf p}$ and a negative relation relation ${\bf n}$.
The result of a temporal-probabilistic join with negation
includes, at each time point, the probability with which a tuple
$\tilde{p}$ of the positive relation ${\bf p}$ matches no tuple in
the negative relation ${\bf n}$ under a predicate $\theta$. 
Firstly, this occurs at time points when either no tuple of
${\bf n}$ has non-zero probability to be valid or no valid tuple
of ${\bf n}$ satisfies the $\theta$-condition. In this case, tuple
$\tilde{p}$ remains {\em unmatched} and the probability of the
output tuple produced is equal to the probability of $\tilde{p}$.

Secondly, the non-existence of a matching tuple for $\tilde{p}$
in ${\bf n}$ occurs when all the valid tuples of ${\bf n}$ that
match $\tilde{p}$ for $\theta$ are false. This case relates
to the probabilistic dimension and thus $\tilde{p}$ is not
disqualified for the output. The output fact is determined by 
$\tilde{p}$ whereas for the computation of the
corresponding probability we need to consider the negating form of
the probabilities for the matching tuples in the negative
relation. In case one of the matching tuples in ${\bf n}$ has
probability equal to $1$, the output tuple has $0$ probability to
be $true$.


\begin{example}
In Fig.~\ref{fig:tpAntiJoin}, the TP anti join of relations
$\mathbf{a}$ and $\mathbf{b}$ of Fig.~\ref{fig:relationsTPDB}
contains, at each time point, the probability that clients
want to visit a location and no hotel is available. Tuple
(\bsq{Ann, ZAK}, $a_1$, [2,4), 0.7) indicates that the tuple
$a_1$ of the positive relation $\mathbf{a}$ remains unmatched
since there is no hotel in {\em ZAK} that has a probability to be
available in the interval [2,4). Tuple (\bsq{Ann, ZAK}, $a_1
\land \lnot (b_3 \lor b_2)$, [5,6), 0.084) corresponds to the
case when the matching tuples of the negative relation
$\mathbf{b}$ are $false$.
\end{example}

\begin{figure}[ht] \centering 
\fontsize{8pt}{11pt}\selectfont
\setlength{\tabcolsep}{4pt}
\centering
\scalebox{1}{
\begin{tabular}{ c c | c | c | c }
\multicolumn{5}{l}{Q = $\mathbf{a} \AJoin^\kat{Tp}_\theta \mathbf{b}$}\\
\hline
$Name$ & $Loc$ & $\lambda$              & $T$         & $p$  \\ \hline
Ann    & ZAK  & $a_1$                  & [2,4)   & 0.7  \\ 
Ann    & ZAK  & $a_1 \land \lnot b_3$  & [4,5) & 0.21\\
Ann    & ZAK  & $a_1 \land \lnot (b_3 \lor b_2)$ & {[5,6)} & 0.084\\
Ann    & ZAK  & $a_1 \land \lnot b_2$  & [6,8) & 0.28\\
Jim    & WEN  & $a_2$				   & [7,10)& 0.8\\
\hline
\end{tabular}}
\caption{$\mathbf{a} \AJoin^\kat{Tp}_\theta \mathbf{b}$ with
$\theta : \mathbf{a}$.Loc = $\mathbf{b}$.Loc ($\mathbf{a}$, $\mathbf{b}$
of Fig.~\ref{fig:relationsTPDB}).}
\label{fig:tpAntiJoin}
\end{figure}

TP outer joins are joins with and without negation. What differs for
outer joins when the temporal and the probabilistic dimension coexist
is that two outcomes might arise at a time point. For example, in
Fig.~\ref{fig:exampleQuery}, the TP left join
$\bf a \ \LJoin^\kat{Tp} \ \ b$ includes, at each time point, cases
when there is a non-zero probability for a tuple in ${\bf a}$ either
to be matched with a tuple in ${\bf b}$ or not based on a predicate
$\theta$. At time point $t=5$, tuple $a_1$ is combined with tuple
$b_3$ producing the output tuples (\bsq{Ann, ZAK, hotel$_2$},
$a_1 \land b_3$, [4,6), 0.49) and (\bsq{Ann, ZAK}, -,
$a_1 \land \lnot b_3)$, [4,5), 0.21) when $b_3$ is $true$ and
\textit{false}, respectively.



%

%% file: 5_windows.tex
\input{5a_windows_definition}


The use of a general $\theta$ condition in TP outer joins and anti
joins requires pairing input tuples that include different facts and
combining multiple input tuples that are valid over an interval and
satisfy $\theta$. For this purpose, we introduce {\bf \em generalized
  lineage-aware temporal windows}, a mechanism created based on two TP
relations ${\bf r}$ and ${\bf s}$, with schema ($F_r$, $F_s$, $T$,
$\mathtt{\lambda_r}$, $\mathtt{\lambda_s}$). $F_r$ and $F_s$ are the
facts included in tuples of relations ${\bf r}$ and ${\bf s}$ over
interval $T$, respectively. $\mathtt{\lambda_r}$ is the disjunction of
the lineage expressions of the tuples of relation ${\bf r}$ that are
valid over $T$, include $F_r$ and satisfy $\theta$.
$\mathtt{\lambda_s}$ is the disjunction of the lineage expressions of
the tuples of relation ${\bf s}$ that are valid over $T$, include
$F_s$ and satisfy $\theta$.



\begin{definition} \label{def:unmatchedIntervals} Let ${\bf r}$
and ${\bf s}$ be TP relations with schema ($F$, $\lambda$, $T$,
$p$) and $\theta$ a condition between the non-temporal attributes
of ${\bf r}$ and ${\bf s}$. The unmatched ${\bf W_\kat{U}}
({\bf r}; {\bf s},\theta)$, overlapping ${\bf W_\kat{O}}
({\bf r}; {\bf s},\theta)$ and negating ${\bf W_\kat{N}}
({\bf r}; {\bf s},\theta)$ windows of ${\bf r}$ with respect to
${\bf s}$ and $\theta$ are defined according to
Table~\ref{tab:windowsDef}. 

\end{definition}

The {\em overlapping windows}
$\mathbf{W_\kat{O}}({\bf r} ; {\bf s}, \theta)$ span a maximal
interval over which a tuple $r$ of ${\bf r}$ overlaps with a tuple $s$
from ${\bf s}$ and the predicate $\theta$ is satisfied. Tuple $r$
includes the fact $F_r$ and has lineage $\lambda_r$ while $F_s$ and
$\lambda_s$ correspond to the fact and lineage of tuple $s$. The
interval of the window that is produced by the pair of tuples $r$ and
$s$ corresponds to the overlap of their interval
($\tilde{w}.T = r.T \cap s.T$). The {\em unmatched windows}
${\bf W_\kat{U}} ({\bf r}; {\bf s}, \theta)$ span over the interval or
a subinterval of a tuple $r$ of ${\bf r}$ during which all tuples of
${\bf s}$ are either not valid or don't satisfy $\theta$
($\lambda^{\mathbf{s}, \theta_{\tilde{w}}}_{t} = \mathtt{null}$).  The
fact $F_r$ and the lineage $\lambda_r$ of an unmatched window are
determined by $r$ while $F_s$ and $\lambda_s$ are set to
$\mathtt{null}$. The negating windows
$\mathbf{W_\kat{N}} ({\bf r}; {\bf s}, \theta)$ of the TP relation
${\bf r}$ with respect to the TP relation ${\bf s}$ are windows during
which a fact is included in a tuple $r$ of ${\bf r}$ as well as in
multiple tuples of ${\bf s}$ that are valid and satisfy the
$\theta$-condition. Negating windows are suitable for producing output
tuples where, for $\theta$, all the tuples of ${\bf s}$ that match a
tuple $r$ of ${\bf r}$ including the fact $F_r$ are \textit{false}, as
described in Section~\ref{sec:properties}. Thus, the fact $F_r$ and
the lineage $\lambda_r$ of the window are determined by $r$, $F_s$ is
set to $\mathtt{null}$ and $\lambda_s$ is the disjunction of the
lineages of all the tuples in ${\bf s}$ that match $r$.

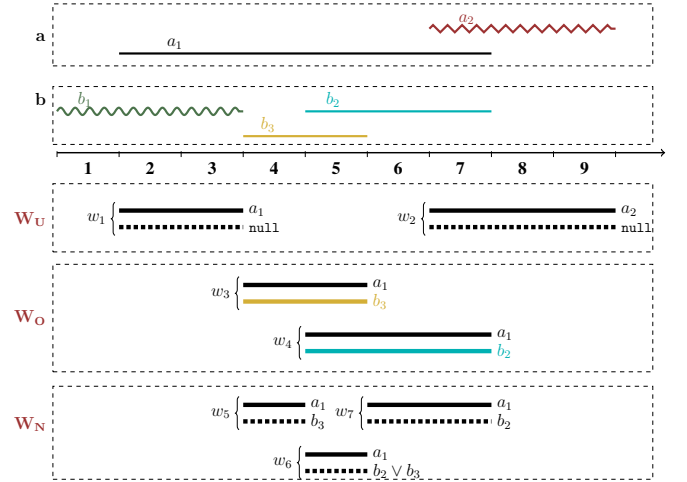
\begin{figure}[!htbp]\centering
\scalebox{0.55} {
\begin{tikzpicture}
\draw (0,0) [->, line width = 1pt]-- coordinate (x axis mid) (14.7,0);
  
\pgfmathsetmacro{\shift}{0.5}
\foreach \j in {1,...,10}{
\pgfmathsetmacro{\divRes}{int(\j/2)}
\pgfmathsetmacro{\modRes}{1-(\j-\divRes*2)}
\pgfmathsetmacro{\xPos}{\j-1+(\divRes-\modRes*\shift)}
\draw (\xPos , 1pt)  --  (\xPos ,-3pt)
node[anchor=north,font=\relsize{2}] {};
}

\pgfmathsetmacro{\shift}{0.5}
\foreach \j in {1,...,9}{
\pgfmathsetmacro{\divRes}{int(\j/2)}
\pgfmathsetmacro{\modRes}{1-(\j-\divRes*2)}
\pgfmathsetmacro{\xPos}{\j-1+(\divRes-\modRes*\shift)}
\draw (\xPos+0.75 , -3pt)
node[anchor=north,font=\relsize{1}] {\bf \j};
}

\pgfmathsetmacro{\winVdist}{0.8}
\pgfmathsetmacro{\xAxisOne}{-0.1}
\pgfmathsetmacro{\xAxisTwo}{14.4}
 
\pgfmathsetmacro{\windowHeight}{1.6}
\pgfmathsetmacro{\yAxis}{0.4}
\draw [line width=1.5,color=c1,snake=coil,segment aspect=0] (0, \yAxis+0.6) -- (4.5,\yAxis+0.6)
node[pos=.15,above=-1pt]{\large $b_1$}; 
\draw [line width=1.5,color=c4] (6, \yAxis+0.6) -- (10.5,\yAxis+0.6)
node[pos=.15,above=-1pt]{\large $b_2$}; 
\draw [line width=1.5,color=c5] (4.5, \yAxis) -- (7.5,\yAxis)
node[pos=.2,above=-1pt]{\large $b_3$}; 
\node at (\xAxisOne-0.3, \yAxis+ 0.9) {\large ${\bf b}$}; 
\pgfmathsetmacro{\yAxis}{\yAxis-0.2}
\draw [dashed] (\xAxisOne,\yAxis) --
               (\xAxisOne,\yAxis + \windowHeight - 0.2) --
               (\xAxisTwo,\yAxis + 	\windowHeight - 0.2) --
               (\xAxisTwo, \yAxis + 0)  -- cycle;

\pgfmathsetmacro{\yAxis}{2.4} 
\pgfmathsetmacro{\windowHeight}{1.5}
\draw [line width=1.5] (1.5, \yAxis) -- (10.5, \yAxis)
node[pos=.15,above=-1pt]{\large $a_1$}; 

\draw [line width=1.5,color=c3, snake=zigzag] (9, \yAxis+0.6) -- (13.5, \yAxis+0.6)
node[pos=.2,above=-1pt]{\large $a_2$}; 
\node at (\xAxisOne-0.3, \yAxis+0.4) {\large ${\bf a}$}; 
\pgfmathsetmacro{\yAxis}{\yAxis-0.3}
\draw [dashed] (\xAxisOne,\yAxis) --
               (\xAxisOne,\yAxis + \windowHeight) --
               (\xAxisTwo,\yAxis + \windowHeight) --
               (\xAxisTwo, \yAxis + 0) -- cycle;
               
\pgfmathsetmacro{\typDist}{0.7}
\pgfmathsetmacro{\windowHeight}{0.7}

\pgfmathsetmacro{\yAxis}{-1.4}
\node[color=c3] at (\xAxisOne-0.55, \yAxis-0.2)
{\large ${\bf W_{U}}$}; 

\draw [line width=3] (1.5,\yAxis) -- (4.5,\yAxis) node[pos=1,right=-1pt] {\large $a_1$};;
\draw [line width=3, dotted] (1.5,\yAxis-0.4) -- (4.5, \yAxis-0.4)
node[pos=1,right=-0.8pt] {$\mathtt{null}$};
\draw[snake=brace] (1.4,\yAxis-0.6) -- (1.4,\yAxis+0.2)
node[pos=0.5, left=2.5pt] {\large $w_1$};


\draw [line width=3] (9,\yAxis) -- (13.5,\yAxis) node[pos=1,right=-1pt] {\large $a_2$};;
\draw [line width=3, dotted] (9,\yAxis-0.4) -- (13.5, \yAxis-0.4)
node[pos=1,right=-0.8pt] {$\mathtt{null}$};
\draw[snake=brace] (8.9,\yAxis-0.6) -- (8.9,\yAxis+0.2)
node[pos=0.5, left=2.5pt] {\large $w_2$};

\pgfmathsetmacro{\yAxis}{-2.4}
\pgfmathsetmacro{\windowHeight}{1.65}
\draw [dashed] (\xAxisOne,\yAxis) --
               (\xAxisOne,\yAxis + \windowHeight) --
               (\xAxisTwo,\yAxis + 	\windowHeight) --
               (\xAxisTwo, \yAxis + 0)  -- cycle;

\pgfmathsetmacro{\num}{1}
\draw [line width=3] (4.5,\yAxis-\winVdist*\num) -- (7.5,\yAxis-\winVdist*\num) node[pos=1,right=-1pt] {\large $a_1$};;
\draw [line width=3, color=c5] (4.5,\yAxis-\winVdist*\num-0.4) -- (7.5,
\yAxis-\winVdist*\num-0.4) node[pos=1,right=-1pt] {\large $b_3$};
\draw[snake=brace] (4.4,\yAxis-\winVdist*\num-0.6) -- (4.4,\yAxis-\winVdist*\num+0.2)
node[pos=0.5, left=2.5pt] {\large $w_3$};
\pgfmathsetmacro{\num}{2.5}
\draw [line width=3] (6,\yAxis-\winVdist*\num) -- (10.5,\yAxis-\winVdist*\num) node[pos=1,right=-1pt] {\large $a_1$};;
\draw [line width=3, color=c4] (6,\yAxis-\winVdist*\num-0.4) -- (10.5,
\yAxis-\winVdist*\num-0.4) node[pos=1,right=-0.8pt] {\large $b_2$};
\draw[snake=brace] (5.9,\yAxis-\winVdist*\num-0.6) -- (5.9,\yAxis-\winVdist*\num+0.2)
node[pos=0.5, left=2.5pt] {\large $w_4$};

\pgfmathsetmacro{\yAxis}{\yAxis-\winVdist*\num-0.9}
\pgfmathsetmacro{\windowHeight}{1.8}
\node[color=c3] at (\xAxisOne-0.55, \yAxis+0.6+0.4*\windowHeight)
{\large $\mathbf{W_{O}}$}; 


\pgfmathsetmacro{\windowHeight}{2.6}
\draw [dashed] (\xAxisOne,\yAxis) --
               (\xAxisOne,\yAxis + \windowHeight) --
               (\xAxisTwo,\yAxis + 	\windowHeight) --
               (\xAxisTwo, \yAxis)  -- cycle;

\pgfmathsetmacro{\num}{1}
\draw [line width=3] (4.5,\yAxis-\winVdist*\num) -- (6,\yAxis-\winVdist*\num) node[pos=1,right=-1pt] {\large $a_1$};;
\draw [line width=3, dotted] (4.5,\yAxis-\winVdist*\num-0.4) -- (6,
\yAxis-\winVdist*\num-0.4) node[pos=1,right=-1pt] {\large $b_3$};
\draw[snake=brace] (4.4,\yAxis-\winVdist*\num-0.6) -- (4.4,\yAxis-\winVdist*\num+0.2)
node[pos=0.5, left=2.5pt] {\large $w_5$};

\pgfmathsetmacro{\num}{2.5}
\draw [line width=3] (6,\yAxis-\winVdist*\num) -- (7.5,\yAxis-\winVdist*\num) node[pos=1,right=-1pt] {\large $a_1$};;
\draw [line width=3, dotted] (6,\yAxis-\winVdist*\num-0.4) -- (7.5,
\yAxis-\winVdist*\num-0.4) node[pos=1,right=-0.8pt] {\large $b_2 \lor b_3$};
\draw[snake=brace] (5.9,\yAxis-\winVdist*\num-0.6) -- (5.9,\yAxis-\winVdist*\num+0.2)
node[pos=0.5, left=2.5pt] {\large $w_6$};

\pgfmathsetmacro{\num}{1}
\draw [line width=3] (7.5,\yAxis-\winVdist*\num) -- (10.5,\yAxis-\winVdist*\num) node[pos=1,right=-1pt] {\large $a_1$};;
\draw [line width=3, dotted] (7.5,\yAxis-\winVdist*\num-0.4) -- (10.5,
\yAxis-\winVdist*\num-0.4) node[pos=1,right=-0.8pt] {\large $b_2$};
\draw[snake=brace] (7.4,\yAxis-\winVdist*\num-0.6) -- (7.4,\yAxis-\winVdist*\num+0.2)
node[pos=0.5, left=2.5pt] {\large $w_7$};

\pgfmathsetmacro{\yAxis}{\yAxis-\winVdist*\num-1.8}
\node[color=c3] at (\xAxisOne-0.55, \yAxis+0.3+0.4*\windowHeight)
{\large $\mathbf{W_{N}}$}; 
\pgfmathsetmacro{\windowHeight}{2.25}
\draw [dashed] (\xAxisOne,\yAxis) --
               (\xAxisOne,\yAxis + \windowHeight) --
               (\xAxisTwo,\yAxis + 	\windowHeight) --
               (\xAxisTwo, \yAxis + 0)  -- cycle;

\end{tikzpicture}}
\caption{All windows of ${\bf a}$ with respect to ${\bf b}$ with $\theta : \mathbf{a}$.Loc = $\mathbf{b}$.Loc}
\label{fig:windowExample_all}
\end{figure}

\begin{example}
In Fig.~\ref{fig:windowExample_all}, the TP relations ${\bf a}$
and ${\bf b}$ of Fig.~\ref{fig:basicTPDBsuper} are illustrated
along with the unmatched, overlapping and negating windows of
${\bf a}$ with respect to ${\bf b}$. Single lines are used for
tuples. Pairs of lines denote windows. Different colors are used
to annotate different facts: black is used for \bsq{Ann, ZAK},
red for \bsq{John, WEN}, green for \bsq{hotel$_3$, SOR}, yellow
for \bsq{hotel$_2$, ZAK}, and blue for \bsq{hotel$_1$, ZAK}.
Wavy lines are used for tuples of an input relation that match
no tuple of the other relation for $\theta$. For the unmatched
window $w_1$ = (\bsq{Ann, ZAK, $\mathtt{null}$}, [2, 4), $a_1$,
$\mathtt{null}$), the straight black line indicates that the fact
$w_1.F_r$ = \bsq{Ann, ZAK} and the lineage $w_1.\lambda_r$ =
$a_1$ match the corresponding attributes of tuple $a_1$. The
dotted line indicates that fact $w_1.F_s$ is $\mathtt{null}$ and
so is $w_1.\lambda_s$. At $t=4$, $a_1$ is still valid whereas
$\lambda^{\mathbf{b}, \theta_{w_1}}_{4} = b_3$, which indicates
that a tuple of ${\bf b}$ starts being valid and thus interval
$[2,4)$ is maximal. The window $w_3$ = (\bsq{Ann, ZAK},
\bsq{hotel$_1$}, [4,6), $a_1$, $b_3$) is an overlapping window.
The blue and a black straight lines for $w_1$ indicate that $F_r$
and $F_s$ of $w_3$ correspond to the facts of tuples $a_1$ and
$b_3$, i.e., tuples that overlap and include the same values for
$Loc$. For the negating window $w_{6}$ = (\bsq{Ann, ZAK},
$\mathtt{null}$, $[5,6)$, $a_1$, $b_3 \lor b_2$), the black
straight line in $w_{6}$ indicates that its fact $F_r$ and its
lineage $\lambda_r$ correspond to the fact and lineage of $a_1$.
The fact $F_s$ is $\mathtt{null}$, illustrated by a dotted line.
Annotated next to this line, the $\lambda_s$ equals the
disjunction of the tuples $b_2$ and $b_3$ that satisfy $\theta$
over the interval $[5,6)$. The interval $[5,6)$ is maximal since
at $t=6$, $b_3$ stops being valid.
\end{example}

An output tuple is formed for each window using the facts $(F_r,
F_s)$ and interval $T$ in their exact form while the output
lineage is formed by combining $\lambda_r$ and $\lambda_s$ with
the proper lineage-concatenation function. According to their
semantics, each set of windows is matched with a unique function:
for {\it overlapping windows} we use the function
$\textit{\textbf{and}}$, for {\it negating windows} we use
$\textit{\textbf{andNot}}$ and for {\it unmatched windows} only
$\lambda_r$ is passed on to the output lineage. For the TP
anti join in Figure~\ref{fig:tpAntiJoin}, the unmatched window
(\bsq{Ann, ZAK}, $\mathtt{null}$, [2,4), $a_1$, $\mathtt{null}$)
is transformed to the output tuple (\bsq{Ann, ZAK}, -, [2,4),
$a_1$) and the negating window (\bsq{Ann, ZAK}, $\mathtt{null}$,
[5,6), $a_1$, $b_3 \lor b_2$) is transformed to the output tuple
(\bsq{Ann, ZAK}, [5,6), $a_1 \land \lnot (b_3 \lor b_2)$). In
Table~\ref{tab:windPerOp}, we include all the window sets
required for each TP join with negation considering that
${\bf W_\kat{O}} ({\bf r}; {\bf s},\theta)$ = 
${\bf W_\kat{O}} ({\bf s}; {\bf r},\theta)$.

\begin{table}[!htpb]
\caption{TP Joins with Negation using Windows}
\label{tab:windPerOp}
\centering
\scalebox{0.9}{
\begin{tabular}{ |  M{0.65cm} | M{1.25cm} | M{1.25cm} | 
					M{1.25cm} | M{1.25cm} | M{1.25cm} |
					@{}m{0cm}@{}}
\hline
\specialrule{1.2pt}{0pt}{0pt}
\textbf{op$^{Tp}$}
& ${\bf W_\kat{U}} ({\bf r}; {\bf s},\theta)$
& ${\bf W_\kat{N}} ({\bf r}; {\bf s},\theta)$
& ${\bf W_\kat{O}} ({\bf r}; {\bf s},\theta)$
& ${\bf W_\kat{U}} ({\bf s}; {\bf r},\theta)$
& ${\bf W_\kat{N}} ({\bf s}; {\bf r},\theta)$ & \\ [0.1cm] 
\hline
${\bf r}\ \AJoin \ {\bf s}$
& \color{c1}{\cmark} & \color{c1}{\cmark} & 
& & & \\ [0.1cm] \hline 
${\bf r}\ \LJoin \ {\bf s}$
& \color{c1}{\cmark} & \color{c1}{\cmark} 
& \color{c1}{\cmark} & & & \\ [0.1cm] \hline 
${\bf r}\ \RJoin \ {\bf s}$
&  &  & \color{c1}{\cmark}
& \color{c1}{\cmark} & \color{c1}{\cmark} & \\ [0.1cm] \hline 
${\bf r}\ \OJoin \ {\bf s}$ 
& \color{c1}{\cmark} & \color{c1}{\cmark}  
& \color{c1}{\cmark} & \color{c1}{\cmark}
& \color{c1}{\cmark} & \\ [0.1cm] \hline 
  
\end{tabular}}
\end{table}

%% file: 5a_windows_definition.tex
\begin{table*}[!htbp] \centering 
\fontsize{8pt}{11pt}\selectfont
\setlength{\tabcolsep}{4pt}
\centering
\caption{}
\label{tab:windowsDef}
\scalebox{1}{
\begin{tabular}{ | c | L{14cm} | @{}m{0cm}@{}}
\hline
\multirow{2}{1.7cm}{Overlapping Windows} 
&  $\tilde{w}\ \in \mathbf{W_\kat{O}}({\bf r}; {\bf s}, \theta) \Longleftrightarrow
\exists r \in \mathbf{r}, \ s \in \mathbf{s} \ (\
  \tilde{w}.F_r = r.F \ \land\ 
  \tilde{w}.F_s = s.F \ \land \ $ & \\
& $\hspace*{2.8cm} \theta \ \land \tilde{w}.\lambda_r \equiv r.\lambda \ \land \ 
 \tilde{w}.\lambda_s \equiv s.\lambda \ \land \
 \tilde{w}.T = r.T \cap s.T\ ) $ & \\
\hline
\multirow{3}{1.7cm}{Unmatched Windows} 
& $\tilde{w}\ \in {\bf W_\kat{U}}({\bf r}; {\bf s}, \theta)
\Longleftrightarrow  \tilde{w}.\lambda_s = \mathtt{null} \ \land
\tilde{w}.F_s = \mathtt{null} \ \land$ & \\
& $\hspace*{2.8cm} \forall t \in \tilde{w}.T \ ( \exists r \in \mathbf{r}\ (\tilde{w}.F_r = r.F \ \land
\tilde{w}.\lambda_r \equiv r.\lambda)\ \land 
\tilde{w}.\lambda_s \equiv \lambda^{\mathbf{s}, \theta_{\tilde{w}}}_{t} \ \land \
\lambda^{\mathbf{s}, \theta_{\tilde{w}}}_{t}=\mathtt{null}) \ \land$ & \\
& $\hspace*{2.8cm} \forall t' \notin \tilde{w}.T \ (
 \nexists r \in \mathbf{r}\ (\tilde{w}.F_r = r.F \land 
 \tilde{w}.\lambda_r \equiv r.\lambda)\lor 
 \tilde{w}.\lambda_s \not\equiv \lambda^{\mathbf{s}, \theta_{\tilde{w}}}_{t'}) $ & \\
\hline
\multirow{3}{1.7cm}{Negating Windows}
& $\tilde{w}\ \in \mathbf{W_\kat{N}}({\bf r}; {\bf s}, \theta) \Longleftrightarrow 
\forall t \in \tilde{w}.T \ ( 
 \exists r \in \mathbf{r}\  (\tilde{w}.F_r = r.F \ \land \
 \tilde{w}.\lambda_r \equiv r.\lambda) \ \land \ $ & \\
& $\hspace*{4.2cm} \tilde{w}.F_s = \mathtt{null} \land\ 
\lambda^{\mathbf{s}, \theta_{\tilde{w}}}_{t} \neq \mathtt{null}
\land \tilde{w}.\lambda_s = \lambda^{\mathbf{s}, \theta_{\tilde{w}}}_{t}\ )\ \land$ & \\
& $\hspace*{2.8cm} \forall t' \notin \tilde{w}.T \ ( \ 
 \nexists r \in \mathbf{r}\ (\tilde{w}.F_r = r.F \ \land \
 \tilde{w}.\lambda_r \equiv r.\lambda) \ \lor 
 \tilde{w}.\lambda_s \not\equiv \lambda^{\mathbf{s}, \theta_{\tilde{w}}}_{t'}\ ) $ & \\
\hline
\end{tabular}}
\end{table*}

%% file: 7_algorithms.tex
In this section, we introduce algorithms to compute 
\emph{generalized lineage-aware temporal windows} and the result
of TP joins with negation. Our Lineage-Aware Window Advancers
(LAWA) for unmatched (LAWA$_\kat{U}$) and negating
(LAWA$_\kat{N}$) windows use overlapping windows as a
computational basis. LAWA$_\kat{U}$ (Algorithm~\ref{algo:lawaUIO})
produces the unmatched windows of ${\bf r}$ with respect to ${\bf
s}$ by identifying the subintervals of ${\bf r}$ during which
there is no overlap or match with a tuple of ${\bf s}$, i.e.,
subintervals that do not correspond to any overlapping window.
Similarly, each of the negating windows of ${\bf r}$ with respect
to ${\bf s}$ spans a subinterval where all tuples of ${\bf s}$
that overlap and match with a tuple $r$ of ${\bf r}$ are false and
thus lineage information from all the overlapping windows that are
valid over this subinterval and involving $r$ must be combined.

LAWA$_\kat{U}$ and LAWA$_\kat{N}$ are sweeping-window
algorithms~\cite{papaioannou2018} that are applied on windows
instead of tuples. They are responsible for forming a set of
windows based on overlapping ones but also for passing the input
windows to the output since they are also necessary for the result
of a TP join with negation. They are operating in an incremental
manner, thus avoiding recomputing the overlapping windows multiple
times.

\input{7a_preprocessing_fig}
\subsection{Overlapping Windows}
\label{sec:lawa_o}

For the computation of overlapping windows of relation ${\bf r}$
with respect to ${\bf s}$, we perform the conventional outer join
${\bf r} \LJoin_{\theta_o \land \theta} {\bf s}$ with
the overlapping predicate $\theta_o: r.T \cap s.T$ and a
condition $\theta$ on the non-temporal attributes, as provided in
the TP join to be computed. The result of ${\bf r}
\LJoin_{\theta_o \land \theta} {\bf s}$ computes a set of windows
enhanced with the time-interval of the tuple of $r$ valid over
each window, and its result has schema: ($F_r$, $\lambda_r$,
$F_s$, $\lambda_s$, $[O_s, O_e)$, $[T_s, T_e)$). ($F_r$, $[T_s,
T_e)$, $\lambda_r$) correspond to the fact, interval and lineage
of a tuple $r$ in ${\bf r}$. Similarly, ($F_s$, $\lambda_s$)
correspond to tuple $s$ in ${\bf s}$. $[O_s, O_e)$ is the
interval during which the tuples $r$ and $s$ overlap. 

\begin{figure}[h]
\fontsize{8pt}{11pt}\selectfont
\setlength{\tabcolsep}{4pt}
\centering
\scalebox{1}{
\begin{tabular}{ c c c c c c | c}
\multicolumn{7}{l}{\hspace{0.7cm}\bf $\textbf{X} $ }\\          
\cline{2-7}
$ $ & $F_r$ & $\lambda_r$ & $F_s$ & $\lambda_s$
& $[O_s, O_e)$  & $[T_s, T_e)$ \\
\cline{2-7}
${\bf x_1}$ & \bsq{Ann, ZAK}       & $a_1$
            & \bsq{hotel$_1$, ZAK} & $b_3$ & [4,6) & [2,8)   \\
${\bf x_2}$ & \bsq{Ann, ZAK}       & $a_1$
            & \bsq{hotel$_2$, ZAK} & $b_2$ & [5,8) & [2,8)   \\
${\bf x_3}$ & \bsq{Jim, WEN}      & $a_2$
			& $\mathtt{null}$ & $\mathtt{null}$ &
			$\mathtt{null}$ & [9,12)  \\
\cline{2-7}
\end{tabular}}
\caption{The result of ${\bf a}\ \LJoin_\kat{\ r.T $\cap$ s.T 
$\land$ {\bf a}.Loc={\bf b}.Loc}\ {\bf b}$.}
\label{fig:lawauo_input}
\end{figure}

The tuples of the join ${\bf r} \LJoin_{\theta_o \land \theta}
{\bf s}$ for which all attributes are not $\mathtt{null}$
constitute the set of overlapping windows
${\bf W_\kat{o}}({\bf r}; {\bf s}, \theta)$. However, the use of
the conventional left join results also in pairs with
$\mathtt{null}$ attributes.

\subsection{Unmatched Windows}

The unmatched windows of a TP relation ${\bf r}$ with respect to
a TP relation ${\bf s}$ and a condition $\theta$ are computed in
two phases. Firstly, the windows in result of ${\bf r}
\LJoin_{\theta_o \land \theta} {\bf s}$ with ($F_s$, $\lambda_s$)
and $[O_s, O_e)$ equal to $\mathtt{null}$ correspond to unmatched 
windows where input tuples of ${\bf r}$ don't overlap or
satisfy $\theta$ with any tuple in ${\bf s}$. The interval
of each such window is equal to the interval $[T_s, T_e)$ of the
tuple of ${\bf r}$.

Secondly, the algorithm LAWA$_\kat{U}$ extends the result
${\bf X}$ of ${\bf r} \LJoin_{\theta_o \land \theta} {\bf s}$
(cf. Fig.~\ref{fig:lawauo_input}) with the remaining unmatched
windows, i.e., the windows that span a {\it subinterval} of a tuple in ${\bf r}$ during which no tuple
in ${\bf s}$ is valid or satisfies $\theta$. For these unmatched
windows to be created, the windows in ${\bf X}$ are grouped
according to the fact $F_r$ and the interval $[T_s, T_e)$ of the
tuple in ${\bf r}$ to which they correspond. Within each group,
the tuples are sorted on the starting point ($Os$) of the 
overlapping intervals and the order of tuples with equal starting
points does not matter. The algorithm performs a sweep of the
interval $[T_s, T_e)$ of each $r$ tuple of ${\bf r}$. It copies
the overlapping windows ($[Os, Oe) \neq \mathtt{null}$) relating
to $r$ to the output. At the same time, given the subintervals
that the overlapping windows span and the initial interval
$[T_s, T_e)$ of $r$, it identifies the subintervals during which
there is no overlap with a tuple in ${\bf s}$, i.e., no
overlapping window, and produces the remaining unmatched windows.

\begin{center}
\begin{algorithm2e}[!h]
\small 
$(\mathtt{prevWindTe, F_r, \lambda_r, wind, PQ, neg}  ) = \mathtt{status}$\;
\BlankLine

\lIf {$\mathtt{wind} = \mathtt{null}$} {\Return $\mathtt{null}$}
\BlankLine

\Do{$\mathtt{windTs} \geq \mathtt{windTe}$}  {

\If {$\mathtt{prevWindTe}=-1$} { \label{line:u_windTsA}
  $\mathtt{windTs} = \mathtt{wind} . T_s$;\
  $\mathtt{F_r} = \mathtt{wind} . F_r$;\ 
  $\mathtt{\lambda_r}$ = $\mathtt{wind} . \lambda_r$\;
}
\lElse {$\mathtt{windTs}$ =  $\mathtt{prevWindTe}$}   \label{line:u_windTsB}
\BlankLine

$\mathtt{\lambda_s}$ = $\mathtt{null}$; \ $\mathtt{F_s}$ = $\mathtt{null}$;\label{line:u_lambda_factA}

\If {$\mathtt{wind} . O_s = \mathtt{windTs}$  \label{line:overlappingA}} {
    $\mathtt{\lambda_s}$ = $\mathtt{wind} . \lambda_s$;
    $\mathtt{F_s}$ = $\mathtt{wind} . F_s$;
} \label{line:u_lambda_factB} 

\BlankLine

\lIf (\tcp*[f]{Case 1}) {$\mathtt{\lambda_s} \neq \mathtt{null}$} {$\mathtt{windTe} = \mathtt{wind} . O_e$ \label{line:u_windTeA}}
\ElseIf{$\mathtt{windTs} = \mathtt{wind} . T_s \wedge \mathtt{wind} . O_s \neq \mathtt{null}$}{$\mathtt{windTe} = \mathtt{wind} . O_s$;\tcp*[f]{Case 2}} 
\ElseIf{$\mathtt{wind} . O_s = \mathtt{null} \vee \mathtt{windTs} = \mathtt{wind} . O_e$} {
  $\mathtt{next}$ = getNextOf($\mathtt{wind} $ )\;
   \If (\tcp*[f]{Case 3}) {$\mathtt{next} \neq \mathtt{null}
    \wedge
   \mathtt{F_r} = \mathtt{next}.F_r$\;}    
   {$\mathtt{windTe} = \mathtt{next}.O_s$}
   \lElse (\tcp*[f]{Case 4,5}) {$\mathtt{windTe} = \mathtt{wind} . T_e$}
    $\mathtt{wind} $  = $\mathtt{next}$ \;

\label{line:u_windTeB}}

\BlankLine
\lIf {$\mathtt{windTe} = \mathtt{wind} . T_e$} {$\mathtt{prevWindTe} = -1$}
\lElse {$\mathtt{prevWindTe}$ = $\mathtt{windTe}$}
\BlankLine

} 

\BlankLine

$\mathtt{out}$ = ($\mathtt{F_r}$, $\mathtt{F_s}$, $\mathtt{windTs}$, $\mathtt{windTe}$, $\mathtt{\lambda_r}$ , $\mathtt{\lambda_s}$) \label{line:window} \;
\BlankLine

$\mathtt{status}$ = ($\mathtt{prevWindTe, F_r, \lambda_r, wind,
PQ, neg}$)\;

\BlankLine
\Return ($\mathtt{window}, \mathtt{status}$)\;
\caption{LAWA$_\kat{U}$($\mathtt{status}$)}
\label{algo:lawaUIO}
\end{algorithm2e}
\vspace*{-0.5cm}
\end{center}

The execution of algorithms LAWA$_\kat{U}$ and LAWA$_\kat{N}$ is
based on a context node ($\mathtt{status}$) with information on
the status of the algorithm: the right boundary of the last
output window ($\mathtt{prevWindTe}$), the fact ($\mathtt{F_r}$)
and the lineage ($\mathtt{\lambda_r}$) of the tuple of ${\bf r}$
that is valid over the output window $[\mathtt{windTs},
\mathtt{windTe})$, and the input window ($\mathtt{wind}$) to be
processed. The tag $\mathtt{neg}$ and the priority queue
$\mathtt{PQ}$ are not used in LAWA$_\kat{U}$. At each call, a
generalized lineage-aware temporal window $\mathtt{out}$ is
returned (Line~\ref{line:window}) as well as the
$\mathtt{status}$ necessary for the next call. Prior to the first
call of LAWA$_\kat{U}$, the first window of ${\bf X}$ is fetched,
$\mathtt{F_r}$ and $\mathtt{\lambda_r}$ are initialized to
$\mathtt{null}$ and $\mathtt{prevWindTe}$ is initialized to $-1$.

{\bf \em Lines~\ref{line:u_windTsA}-\ref{line:u_windTsB}:} 
Initially, the left boundary $\mathtt{windTs}$ of the new window
as well as the fact and the lineage of the valid tuple of
${\bf r}$ are determined. If a new group is being processed
($\mathtt{prevWindTe}$ = $-1$), $\mathtt{windTs}$
is determined by the starting point of the first window
$\mathtt{wind}$ of the new group. In this case, the fact
$\mathtt{F_r}$ and the lineage $\lambda_r$  of the valid tuple of
${\bf r}$ are also extracted from $\mathtt{wind}$. If the
processing of a group continues, the interval of the new window
is adjacent to the previous one, with $\mathtt{windTs}$ =
$\mathtt{prevWindTe}$ while $\mathtt{F_r}$ and $\lambda_r$ remain
unchanged.


{\bf\em
Lines~\ref{line:u_lambda_factA}-\ref{line:u_lambda_factB}:}
In order to determine the fact and the lineage of the tuple
of ${\bf s}$ valid over the output window, we check if the 
starting point $\mathtt{windTs}$ of the window matches the
starting point $O_s$ of an overlapping window in ${\bf X}$. If 
satisfied, this condition (Line~\ref{line:overlappingA})
indicates that there is a tuple of ${\bf s}$ valid over the
window and thus the fact $\mathtt{F_s}$ and lineage $\lambda_s$
equal the corresponding attributes of $\mathtt{wind}$. Otherwise,
they are set to $\mathtt{null}$.

\input{7a_fig_LAWAu_cases}

{\bf \em Lines~\ref{line:u_windTeA}-\ref{line:u_windTeB}:} 
The right boundary $\mathtt{windTe}$ of $\mathtt{out}$ is 
determined based on whether it is an overlapping or an  
unmatched one. All the cases are annotated in the algorithm and
illustrated in Figure~\ref{fig:algoCases}. If
$\mathtt{out}$ is an overlapping window (Case 1), i.e.,
$\mathtt{\lambda_s} \neq \mathtt{null}$, its interval corresponds
to the overlapping interval in $\mathtt{wind}$ and thus,
$\mathtt{windTe}$ is set to $\mathtt{wind.Oe}$. If the output
window is an unmatched window, three different cases are
considered based on the position of $\mathtt{windTs}$ with
respect to $[\mathtt{wind}.O_s, \mathtt{wind}.O_e)$. If the
starting point $\mathtt{windTs}$ coincides with the starting
point of the valid tuple of ${\bf r}$ ($\mathtt{windTs} =
\mathtt{wind}. T_s$) and the starting point of the overlapping 
window $\mathtt{wind}$ succeds (Case 2), $\mathtt{windTe}$ is set
to the starting point of $\mathtt{wind}$. If the starting point
of the output window coincides with the ending point of the
overlapping window $\mathtt{wind}$ (Case 3), the upcoming window
$\mathtt{next}$ is fetched. If $\mathtt{next}$ is in the
same group as $\mathtt{wind}$, $\mathtt{out}$ is positioned
between two overlapping windows and thus $\mathtt{windTe} =
\mathtt{next}.Os$. However, if $\mathtt{next}$ belongs to a new
group, $\mathtt{wind}$ is positioned at the end of the 
interval of a valid tuple of ${\bf r}$ (Case 4). Thus
$\mathtt{windTe} = \mathtt{wind}.Te$ and the sweeping progresses
to window $\mathtt{next}$. The same assignment takes place if
$\mathtt{wind}$ is one of the unmatched windows produced by the
conventional left outer join (Case 5).

\input{7a_fig_LAWAu}

\begin{example}
In Fig.~\ref{fig:algoUOrun}, we illustrate two calls of
LAWA$_{U}$ when applied on relation ${\bf X}$ of
Fig~\ref{fig:lawauo_input} and more specifically on the group of
windows with the fact $\mathtt{F_r} = $\bsq{Ann, ZAK}. The single
blank line corresponds to tuple $a_1$, the tuple of the left
relation ${\bf a}$ valid over all windows of the group. The
window $\mathtt{wind} = x_1$ is the first to be processed. In the
first call of LAWA$_{U}$, illustrated at the bottom of the
figure, the processing of a new group starts and
$\mathtt{windTs}$, $\mathtt{F_r}$ and $\mathtt{\lambda_r}$ are
initialized to the starting point, fact and lineage of $a_1$,
respectively. No overlapping window of the same group starts at
$\mathtt{windTs} = 2$ and thus, $\mathtt{F_s}$ and
$\mathtt{\lambda_s}$ are set to $\mathtt{null}$. According to
Case 2, $\mathtt{windTe}$ is set to $\mathtt{wind}.O_s$. In the
second call of LAWA, the same group is processed and
$\mathtt{out}$ will be adjacent to the previous output 
window. Since $\mathtt{windTs}$ equals the starting point of
the overlapping window $x_1$, the facts, lineages and
intervals of the output window are fetched from $x_1$. The ending
point $\mathtt{windTe}$ of $\mathtt{out}$ is set according to Case
1.
\end{example}

\subsection{Negating Windows}
\label{sec:lawan}

LAWA$_\kat{N}$ extends the result ${\bf Y}$ of LAWA$_\kat{U}$ with
the negating windows. ${\bf Y}$ consists of windows ordered
by the fact of ${\bf r}$ ($F_r$) as well as by their starting
point ($Ts$). LAWA$_\kat{N}$ sweeps over
${\bf Y}$ and copies all the unmatched and overlapping windows
to the output. When a group of overlapping windows with the same
fact $F_r$ is encountered, negating windows are created. The
intervals of these windows are subintervals of the group of
overlapping windows.



\begin{figure}[!ht]
\fontsize{8pt}{11pt}\selectfont
\setlength{\tabcolsep}{2pt}
\centering
\scalebox{1}{
\begin{tabular}{r c c|c c| c}
\multicolumn{6}{l}{\hspace{0.7cm}\bf $\textbf{Y} $ }\\          
\cline{2-6}
$ $  & {$F_r$}  & {$F_s$}
     & $\lambda_r$ & $\lambda_s$ & $T = [T_s, T_e)$ \\ \cline{2-6}
$\bf y_1$ & \bsq{Ann, ZAK} &  $\mathtt{null}$
          & $a_1$          &  $\mathtt{null}$    & [2,4)   \\
$\bf y_2$ & \bsq{Ann, ZAK} & \bsq{hotel$_1$, ZAK}
		  & $a_1$ & $b_3$ & [4,6)   \\
$\bf y_3$ & \bsq{Ann, ZAK} & \bsq{hotel$_1$, ZAK}
          & $a_1$ & $b_2$ & [5,8)   \\
$\bf y_4$ & \bsq{Jim, WEN} & $\mathtt{null}$
		  & $a_2$ & $\mathtt{null}$ & [9,12)   \\
\cline{2-6}
\end{tabular}}
\caption{The input of LAWA$_{N}$}
\label{fig:lawan_input}
\end{figure}

The execution of LAWA$_\kat{N}$ is also based on the context node
$\mathtt{status}$. The tag $\mathtt{neg}$ indicates if a negating
window will be produced. The priority queue $\mathtt{PQ}$
includes $(t, \lambda)$ pairs that indicate the time point $t$
after which the tuple of the right relation with lineage
$\lambda$ stops being valid.

{\bf \em Lines~\ref{line:startNA}-\ref{line:startNB}:} In the
first call of the algorithm ($\mathtt{firstCall}$), the first
tuple of ${\bf Y}$ is fetched, the priority queue $\mathtt{PQ}$ is
initialized (pointer to $\mathtt{null}$), $\mathtt{prevWindTe}$ is
set to $-1$ and $\mathtt{neg}$ to $false$. Since negating windows
are created based on the overlapping windows, whenever a group of
overlapping windows with the same $F_r$ starts, the output fact
$\mathtt{F_r}$, the output lineage $\mathtt{\lambda_r}$ and the
starting point $\mathtt{prevWindTe}$ of the output windows are
updated to the values of the first tuple of this group for $F_r$,
$\mathtt{\lambda_r}$ and $T_s$ respectively.

\begin{center}
\begin{algorithm2e}[ht]
\small 
$(\mathtt{prevWindTe, F_r, \lambda_r, wind, PQ, neg}  ) = \mathtt{status}$\label{line:startNA}\;
\BlankLine

\lIf {$\mathtt{wind}=\mathtt{null} \land isPQempty()$}
{\Return ($\mathtt{null}$, $\mathtt{null}$)}

\If {$\mathtt{firstCall}$} {
  $\mathtt{PQ}$ = initializePQ();  $\mathtt{prevWindTe} = -1$; \ $\mathtt{neg} = false$\; 
}
\BlankLine


 
\If {$\mathtt{prevWindTe}=-1 \land \mathtt{wind.\lambda_r} \neq \mathtt{null}$} {
	$\mathtt{F_r} = \mathtt{wind}.F_r$;\
	$\mathtt{\lambda_r}$ = $\mathtt{wind}$.$\lambda_r$; \
	$\mathtt{prevWindTe} = \mathtt{wind}.T_s$;\label{line:startNB}\ 
}

\BlankLine

\While{$\mathtt{out} = \mathtt{null}$} {
\If{$\mathtt{neg} = false$\label{line:negCheckA}} {
   $\mathtt{out} = \mathtt{wind}$\label{line:copy}\;
   \lIf {$\mathtt{wind} . F_s = \mathtt{null}$} {	
     $\mathtt{wind}$  = getNextTuple()}
   \lElse{
	$\mathtt{neg} = true$\label{line:negFalse};
    addToPQ($\mathtt{wind}. T_e, \mathtt{wind} . \lambda_s$)\label{line:queue}} 
  
}
\ElseIf {$\mathtt{wind} . F_r  = \mathtt{F_r} \land 
		  \mathtt{wind} . Ts \leq \mathtt{prevWindTe}$} {
  $\mathtt{wind} $  = getNextTuple() \label{line:next}\;
\label{line:negCheckB}}

\BlankLine
\If {$\mathtt{out=null} \land \mathtt{wind} . F_r = \mathtt{F}$
\label{line:negA}}{

	\If{$\mathtt{wind} . T_s > \mathtt{prevWindTe}$} {
      $\mathtt{windTe}$ = tForTopOfPQ()\;
      \If{$\mathtt{wind} . T_s < \mathtt{windTe}$} {
        $\mathtt{windTe} = \mathtt{wind} . T_s$\; 
        $\lambda_s$ = disjunctLineages($\mathtt{windTe}$)\;
      
      	$\mathtt{out} = (\mathtt{\mathtt{F_r}}, -,
					   [\mathtt{prevWindTe}, \mathtt{windTe}),
					   \lambda_r, \lambda_s)$\;
    	$\mathtt{prevWindTe}$ = $\mathtt{windTe}$\;
    	$\mathtt{neg} = false$\;
   	}
  } \lElseIf {$\mathtt{wind} . T_s = \mathtt{prevWindTe}$} {
    	$\mathtt{neg} = false$\label{line:special}}
}
\BlankLine
\ElseIf {$\mathtt{out=null} \land (\lnot$ isPQempty()$)$
\label{line:PQcheck}} {
  $\mathtt{windTe}$ = tForTopOfPQ()\label{line:windTeFromQ}\;
  $\lambda_s$ = disjunctLineages($\mathtt{windTe}$)\;
  $\mathtt{out} = (\mathtt{F_r}, -,
	 			   [\mathtt{prevWindTe}, \mathtt{windTe}),
				   \lambda_r, \lambda_s)$\;
  $\mathtt{prevWindTe}$ = $\mathtt{windTe}$;\ removeTopOfPQ()\; 
\label{line:negB}  }

\BlankLine

%
}\label{algo:whileEnd}
\BlankLine

\lIf {isPQempty()} {
$\mathtt{prevWindTe} = -1$; \ $\mathtt{neg} = false$}

\BlankLine
$\mathtt{status}$ = $(\mathtt{prevWindTe, F_r, \lambda_r, wind,
PQ, neg}  )$\;
\BlankLine
\Return ($\mathtt{out}, \mathtt{status}$)\;
\caption{LAWA$_{N}$($\mathtt{status}$)}
\label{algo:lawaN}
\end{algorithm2e}
\vspace*{-0.5cm}
\end{center}

{\bf \em Lines~\ref{line:negCheckA}-\ref{line:negCheckB}:} 
LAWA$_\kat{N}$ outputs an unmatched, overlapping or negating  
window according to $\mathtt{neg}$. When $\mathtt{neg}$ is $false$
(Line~\ref{line:negCheckA}), the unmatched or overlapping window
$\mathtt{wind}$ is copied to the output as is
(Line~ ref{line:copy}). If $\mathtt{wind}$ corresponds to an
unmatched window ($\mathtt{wind}.F_s$ = $\mathtt{null}$), we
proceed to the next window. However, if it corresponds to an
overlapping window, the creation of a negating window follows and
$\mathtt{neg}$ is set to $true$ (Line~\ref{line:negFalse}). In
this case, we add to $\mathtt{PQ}$ the pair ($\mathtt{wind}.T_e$,
$\mathtt{wind}.\lambda_s$), with the ending point and the lineage
of the valid tuple in the relation ${\bf s}$ as recorded in
$\mathtt{wind}$.

When $\mathtt{neg}$ is $true$, the creation of a negating window 
follows. If the same group is processed and the starting point of
$\mathtt{out}$ ($\mathtt{prevWindTe}$) is equal to the starting
point of $\mathtt{wind}$, the next window is fetched
(Line~\ref{line:next}) for two reasons. Firstly, if the next
window of ${\bf Y}$ is an overlapping window of the same group
and starts at $\mathtt{prevWindTe}$, the lineage of the tuple of
relation ${\bf s}$ valid over this input window needs to be
considered for $\mathtt{\lambda_s}$. Secondly, if the next window
belongs to the same group, its starting point should be
considered as a potential ending point of $\mathtt{out}$. 

{\bf \em Lines~\ref{line:negA}-\ref{line:special}:} The
output negating window is finalized by determining its ending
point $\mathtt{windTe}$ and lineage $\mathtt{\lambda_s}$. The 
lineage $\mathtt{\lambda_s}$ is always determined by
disjuncting the lineage expressions of the pairs $(t,
\lambda)$ in the priority queue with $t$ smaller than
$\mathtt{windTe}$. Thus, $\mathtt{\lambda_s}$ correspond to
the dinjuction of the tuples of the relation ${\bf s}$ valid
over the output interval $[\mathtt{prevWindTe}, \mathtt{windTe})$.
To determine the ending point $\mathtt{windTe}$ of the window, we
first check if the upcoming window $\mathtt{wind}$ of ${\bf Y}$
includes the same fact $\mathtt{F_r}$ as $\mathtt{out}$. If this
is the case, $\mathtt{windTe}$ is the minimum between the time
point of the top pair in the queue, i.e., the smallest ending
point of valid tuples in relation ${\bf s}$, and the starting
point of the upcoming window of ${\bf Y}$. Therefore, a window is
created  when there is a change in the tuples of relation
${\bf s}$ that are valid either because a tuple ends or a new
tuple begins. After $\mathtt{out}$ is formed, the starting point
$\mathtt{prevWindTe}$ of the next negating window is set to
$\mathtt{windTe}$. $\mathtt{neg}$ is set to $false$ so that
the window $\mathtt{wind}$ is copied to the output.

A special case occurs when the starting point of the upcoming 
window is equal to the starting point of the output window
(Line~\ref{line:special}). This means that there exists a valid
tuple in the reference relation ${\bf s}$ that needs to be
considered for the output window and thus its finalization is
postponed. The upcoming window, either overlapping or unmatched,
has to be first copied to the output so we set $\mathtt{neg}$ 
back to false.

{\bf \em Lines~\ref{line:PQcheck}-\ref{line:negB}:} 
If there are more overlapping windows in $\mathtt{PQ}$ that
end before the upcoming window $\mathtt{wind}$ starts, regardless
of whether $\mathtt{wind}$ belongs in the same or a different 
group, the ending point of the new negating window is equal to the
ending point of the pair on top of the priority queue
(Line~\ref{line:windTeFromQ}. The starting point of the next
negating window is set to $\mathtt{windTe}$ indicating that the
sweeping until this time point has been completed. As a result, all
the nodes in $\mathtt{PQ}$ correspond to windows whose ending
point is equal to $\mathtt{windTe}$ have already been considered
and need to be removed.

\input{7a_fig_LAWAn}

\begin{example}
In Fig.~\ref{fig:algoNrun}, we focus on the group with
$\mathtt{F_r}$=\bsq{Ann, ZAK} and we illustrate all six calls
of LAWA$_\kat{N}$ on the corresponding windows of the result
$\bf Y$ of LAWA$_\kat{U}$ (Fig.\ref{fig:lawan_input}), when
applied on the relations ${\bf a}$ and ${\bf b}$ of
Fig.\ref{fig:relationsTPDB}. Red color is used for windows copied
to the output whereas green is used for the negating windows.
In the first two calls of LAWA$_\kat{N}$, windows $y_1$ and $y_2$
are copied to the output. $y_2$ is the first overlapping window 
after a series of unmatched ones. After $\mathtt{out}=y_2$,
$\mathtt{neg}$ is set to $true$ and the sweeping for negating
tuples starts from $\mathtt{prevWindTe} = y_2.Ts = 4$ with
$\mathtt{F_\lambda}$ = \bsq{Ann,ZAK} and $\mathtt{\lambda_r} =
a_1$. Window $y_2$ is followed by another overlapping window
($y_3$) that starts before the ending point of $y_2$, recorded in
the top node of the priority queue. Consequently,
$\mathtt{windTe} = y_4.Ts = 5$ and the negating window (\bsq{Ann,
ZAK}, $\mathtt{null}$, [4, 5), $a_1$, $b_3$) is produced.
$\mathtt{neg}$ is set false and window $y_3$ is then copied to
the output. Since there are no more overlapping windows to be processed, the upcoming negating windows are adjacent to
each other and their ending points are derived from the nodes of
$\mathtt{PQ}$.
\end{example}


\subsection{TP Join Algorithms}
\label{sec:basicTPn}

In this subsection we introduce the algorithm
\emph{NegationJoins}($\mathbf{r}$, $\mathbf{s}$, $\theta$,
$\mathtt{op}$) that computes the result of the TP outer join
or anti join $\mathtt{op}$ on the input TP relations $\mathbf{r}$
and $\mathbf{s}$ and the predicate $\theta$. In contrast to
previous works in either temporal or probabilistic databases,
this algorithms involves no tuple replication. Instead, it allows
for a pipelined calculation of the result and thus enables its
smooth integration in the kernel of a DBMS. 


\begin{center}
\begin{algorithm2e}[!htbp]
\small 
$\mathbf{w_{init}}$ = \emph{leftJoin}($\mathbf{r}$,
					$\mathbf{s}$,$\theta \land \theta_{o}$)\;
\emph{sort}($\mathbf{w_{init}}$\{$F_L,\mathtt{O_s}$\}) \label{line:sort} \;
  
\BlankLine
$\mathtt{status} = (\mathtt{-1, null, null,
				 fetchWind(\mathbf{w_{init}}),null,false})$\;

\While{$\mathtt{status} \neq \mathtt{null}$}{  \label{line:terminA}
$(\mathtt{w}, \mathtt{status})$ =
				\emph{LAWA$_u$}($\mathtt{status}$)\;
$\mathbf{w_{uo}}$ = $\mathbf{w_{uo}} \cup \{\mathtt{w}\}$\;
\label{line:candidates}
}

\BlankLine
$\mathtt{status} = (\mathtt{-1, null, null,
				 fetchWind(\mathbf{w_{uo}}),null,false})$\;
\While{$\mathtt{status} \neq \mathtt{null}$}{
    \label{line:terminB}
$(\mathtt{w}, \mathtt{status})$ =
					\emph{LAWA$_n$}($\mathtt{status}$)\;

\If{$\mathtt{w.\lambda_s} = \mathtt{null}$ $\land$
    $\mathtt{w.F_s} = \mathtt{null}$} {
\label{line:filtering}
$\mathtt{o}$ = $\mathtt{o}$ $\cup$
	          \{($\mathtt{w.F_r}$, $\mathtt{w.F_s}$,
	             $\mathtt{w.\lambda_r}$,
          [$\mathtt{w.winTs}$, $\mathtt{w.winTe}$))\}\;
}

\ElseIf{$\mathtt{w.\lambda_s} \neq \mathtt{null}$ $\land$
        $\mathtt{w.F_s} = \mathtt{null}$} {
\label{line:filtering}
$\mathtt{\lambda}$ = \textbf{andNot}($\mathtt{w.\lambda_r}$, 	
									$\mathtt{w.\lambda_s}$)\;
$\mathtt{o}$ = $\mathtt{o}$ $\cup$
			  \{($\mathtt{w.F_r}$, $\mathtt{w.F_s}$,
			  	 $\mathtt{\lambda}$,
			  	[$\mathtt{w.winTs}$, $\mathtt{w.winTe}$))\}\;
}

\ElseIf{$\mathtt{op} \neq \AJoin$} {
\label{line:filtering}
$\mathtt{\lambda}$ = \textbf{and}($\mathtt{w.\lambda_r}$,
								  $\mathtt{w.\lambda_s}$)\;
$\mathtt{o}$ = $\mathtt{o}$ $\cup$ 
			  \{($\mathtt{w.F_r}$, $\mathtt{w.F_s}$,
			     $\mathtt{\lambda}$,
	        [$\mathtt{w.winTs}$, $\mathtt{w.winTe}$))\}\;
}}

\BlankLine
\lIf {$\mathtt{op}$ = $\OJoin$} {
$\mathtt{o}$ = $\mathtt{o}$ $\cup$
			   \emph{NegatingJoins}(${\bf s}$, ${\bf r}$,
		  	   $\theta$, $\mathtt{\AJoin}$)
}

\Return $\mathtt{o}$\;
\caption{\emph{NegationJoins}($\mathbf{r}$, $\mathbf{s}$,
							  $\theta$, $\mathtt{op}$)}
\label{algo:negatingJoins}
\end{algorithm2e}
\vspace*{-0.5cm}
\end{center}

Initially, the set $\mathbf{w_{init}}$ includes the overlapping
windows of $\mathbf{r}$ and $\mathbf{s}$ and a subset of the
unmatched windows (Section~\ref{sec:lawa_o}). The windows in
$\mathbf{w_{init}}$ are sorted based on the fact $F_r$ and the
starting point $\mathtt{Ts}$ (Line~\ref{line:sort}) of the tuple
of the positive relation from which they have been produced. As
long as the terminating condition (Line~\ref{line:terminA})
is satisfied, LAWA$_u$ passes through all start and end points of
the windows in $\mathbf{w_{init}}$ in a smaller-to-larger fashion
and expands the set with the unmatched windows
(Line~\ref{line:candidates}) that hadn't been created yet. 
Similarly, LAWA$_n$ sweeps the windows of the set
$\mathbf{w_{uo}}$ and extends it with the negating windows of
$\mathbf{r}$ and $\mathbf{s}$.

Each window $\mathtt{w}$ that LAWA$_n$ produces is not further
swept and it can be transformed to an output tuple for the result
of the TP join. A lineage-based filter is directly applied to
determine if $\mathtt{w}$ is unmatched ($\mathtt{w.\lambda_s} =
\mathtt{null}$ $\land$ $\mathtt{w.F_s} = \mathtt{null}$), negating
($\mathtt{w.\lambda_s} \neq \mathtt{null}$ $\land$
$\mathtt{w.F_s} = \mathtt{null}$) or overlapping. If the join
performed is a TP anti join ($\AJoin^\kat{TP}$), then the
overlapping windows are filtered out and are not included in the
final result. If it is a full outer join, the unmatched and
negating windows of $\mathbf{s}$ using $\mathbf{r}$ as a reference
need to be included and thus the \emph{NegationJoins} algorithm
needs to be called again with reversed arguments, same predicate
and anti join as the operation to be performed so that the
overlapping windows are not copied again to the output. Finally,
every window is finalized into an output tuple using the
lineage-concatenating function that corresponds to set of windows
to which it belongs. In the case of a TP anti join, $\mathtt{F_r}$
is the only fact included in the output tuples.

%% file: 7a_fig_LAWAu_cases.tex
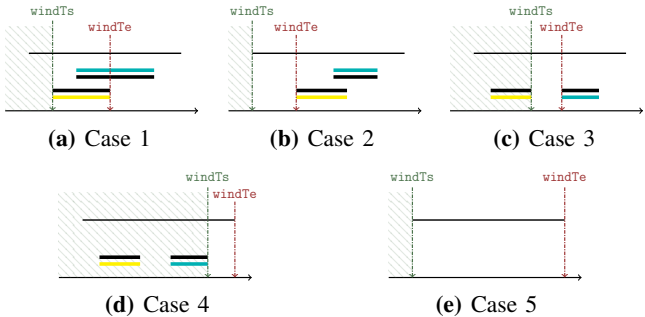
\begin{figure}[!htbp]
\centering
\tikzset{cross/.style={cross out, draw=blue, thick, minimum size=7pt, inner sep=0pt, outer sep=0pt},
cross/.default={1pt}}

\begin{subfigure}[b]{0.32\linewidth} \centering
\scalebox{0.45} {
\begin{tikzpicture}   

\pgfmathsetmacro{\xAxisOne}{-0.2}
\pgfmathsetmacro{\xAxisTwo}{13.7}
\pgfmathsetmacro{\relationwindowHeight}{1}
\pgfmathsetmacro{\gap}{1.8}
\pgfmathsetmacro{\windowHeight}{2.7}

\pgfmathsetmacro{\yAxis}{0}
\pgfmathsetmacro{\yPosA}{\yAxis  + 0.6}
\pgfmathsetmacro{\yPosB}{\yAxis  + 2*0.7 + 0.3}
\pgfmathsetmacro{\yPosC}{\yAxis  + 3*0.7 + 0.3}

\draw (-0.7,\yAxis) [->, line width = 1pt]
    -- coordinate (x axis mid) (5,\yAxis);
  
\pgfmathsetmacro{\winTs}{-0.7}
\pgfmathsetmacro{\winTe}{0.7}
\draw [ultra thick, draw=none, pattern color=c1, pattern=north west lines, opacity=0.4]
(\winTs,\yAxis) -- (\winTs,\yAxis + \windowHeight - 0.2) -- (\winTe,\yAxis + \windowHeight - 0.2) -- (\winTe, \yAxis + 0)  -- cycle;

\draw [line width=1.2] (0, \yPosB) -- (4.5, \yPosB);

\draw [line width=3] (0.7,\yAxis+0.6) -- (2.4,\yAxis+0.6)
 node[pos=1,right=-1pt] {};
\draw [line width=3, color=yellow] (0.7,\yAxis+0.4) -- (2.4,\yAxis+0.4)
node[pos=1,right=-1pt] {};

\draw [line width=3] (1.4,\yAxis+1) -- (3.7,\yAxis+1)
 node[pos=1,right=-1pt] {};
\draw [line width=3, color=c4] (1.4,\yAxis+1.2) -- (3.7,\yAxis+1.2)
node[pos=1,right=-1pt] {};

\pgfmathsetmacro{\winTs}{0.7}
\draw [<-,line width=1,densely dashdotted,c1]
(\winTs, \yAxis) -- (\winTs,\yAxis +\windowHeight)
node[pos=1.1]
{\large {\color{c1} $\mathtt{windTs}$}};

\pgfmathsetmacro{\winTe}{2.4}
\draw [<-,line width=1,densely dashdotted,c3]
(\winTe, \yAxis) -- (\winTe,\yAxis +\windowHeight-0.45)
node[pos=1.1]
{\large {\color{c3} $\mathtt{windTe}$}};

\end{tikzpicture}}
\caption{Case 1}
\label{fig:noupc}
\end{subfigure}
\begin{subfigure}[b]{0.32\linewidth} \centering
\scalebox{0.45} {
\begin{tikzpicture}   

\pgfmathsetmacro{\xAxisOne}{-0.2}
\pgfmathsetmacro{\xAxisTwo}{13.7}
\pgfmathsetmacro{\relationwindowHeight}{1}
\pgfmathsetmacro{\gap}{1.8}
\pgfmathsetmacro{\windowHeight}{2.7}

\pgfmathsetmacro{\yAxis}{0}
\pgfmathsetmacro{\yPosA}{\yAxis  + 0.6}
\pgfmathsetmacro{\yPosB}{\yAxis  + 2*0.7 + 0.3}
\pgfmathsetmacro{\yPosC}{\yAxis  + 3*0.7 + 0.3}

 \draw (-0.7,\yAxis) [->, line width = 1pt]
    -- coordinate (x axis mid) (5,\yAxis);

\pgfmathsetmacro{\winTs}{-0.7}
\pgfmathsetmacro{\winTe}{0}
\draw [ultra thick, draw=none, pattern color=c1, pattern=north west lines, opacity=0.4]
(\winTs,\yAxis) -- (\winTs,\yAxis + \windowHeight - 0.2) -- (\winTe,\yAxis + \windowHeight - 0.2) -- (\winTe, \yAxis + 0)  -- cycle;

\draw [line width=1.2] (0, \yPosB) -- (4.5, \yPosB);

\draw [line width=3] (1.3,\yAxis+0.6) -- (2.8,\yAxis+0.6)
 node[pos=1,right=-1pt] {};
\draw [line width=3, color=yellow] (1.3,\yAxis+0.4) -- (2.8,\yAxis+0.4)
node[pos=1,right=-1pt] {};

\draw [line width=3] (2.4,\yAxis+1) -- (3.7,\yAxis+1)
 node[pos=1,right=-1pt] {};
\draw [line width=3, color=c4] (2.4,\yAxis+1.2) -- (3.7,\yAxis+1.2)
node[pos=1,right=-1pt] {};

\pgfmathsetmacro{\winTs}{0}
\draw [<-,line width=1,densely dashdotted,c1]
(\winTs, \yAxis) -- (\winTs,\yAxis +\windowHeight)
node[pos=1.1]
{\large {\color{c1} $\mathtt{windTs}$}};

\pgfmathsetmacro{\winTe}{1.3}
\draw [<-,line width=1,densely dashdotted,c3]
(\winTe, \yAxis) -- (\winTe,\yAxis +\windowHeight-0.45)
node[pos=1.1]
{\large {\color{c3} $\mathtt{windTe}$}};
 
\end{tikzpicture}}
\caption{Case 2}
\end{subfigure}
\begin{subfigure}[b]{0.32\linewidth} \centering
\scalebox{0.45} {
\begin{tikzpicture}   

\pgfmathsetmacro{\xAxisOne}{-0.2}
\pgfmathsetmacro{\xAxisTwo}{13.7}
\pgfmathsetmacro{\relationwindowHeight}{1}
\pgfmathsetmacro{\gap}{1.8}
\pgfmathsetmacro{\windowHeight}{2.7}

\pgfmathsetmacro{\yAxis}{0}
\pgfmathsetmacro{\yPosA}{\yAxis  + 0.6}
\pgfmathsetmacro{\yPosB}{\yAxis  + 2*0.7 + 0.3}
\pgfmathsetmacro{\yPosC}{\yAxis  + 3*0.7 + 0.3}

 \draw (-0.7,\yAxis) [->, line width = 1pt]
    -- coordinate (x axis mid) (5,\yAxis);

\pgfmathsetmacro{\winTs}{-0.7}
\pgfmathsetmacro{\winTe}{1.7}
\draw [ultra thick, draw=none, pattern color=c1, pattern=north west lines, opacity=0.4]
(\winTs,\yAxis) -- (\winTs,\yAxis + \windowHeight - 0.2) -- (\winTe,\yAxis + \windowHeight - 0.2) -- (\winTe, \yAxis + 0)  -- cycle;

\draw [line width=1.2] (0, \yPosB) -- (4.5, \yPosB);

\draw [line width=3] (0.5,\yAxis+0.6) -- (1.7,\yAxis+0.6)
 node[pos=1,right=-1pt] {};
\draw [line width=3, color=yellow] (0.5,\yAxis+0.4) -- (1.7,\yAxis+0.4)
node[pos=1,right=-1pt] {};

\draw [line width=3] (2.6,\yAxis+0.6) -- (3.7,\yAxis+0.6)
 node[pos=1,right=-1pt] {};
\draw [line width=3, color=c4] (2.6,\yAxis+0.4) -- (3.7,\yAxis+0.4)
node[pos=1,right=-1pt] {};

\pgfmathsetmacro{\winTs}{\winTe}
\draw [<-,line width=1,densely dashdotted,c1]
(\winTs, \yAxis) -- (\winTs,\yAxis +\windowHeight)
node[pos=1.1]
{\large {\color{c1} $\mathtt{windTs}$}};

\pgfmathsetmacro{\winTe}{2.6}
\draw [<-,line width=1,densely dashdotted,c3]
(\winTe, \yAxis) -- (\winTe,\yAxis +\windowHeight-0.45)
node[pos=1.1]
{\large {\color{c3} $\mathtt{windTe}$}};
 
\end{tikzpicture}}
\caption{Case 3}
\end{subfigure}

\vspace*{0.2cm}

\begin{subfigure}[b]{0.48\linewidth} \centering
\scalebox{0.45} {
\begin{tikzpicture}   

\pgfmathsetmacro{\xAxisOne}{-0.2}
\pgfmathsetmacro{\xAxisTwo}{13.7}
\pgfmathsetmacro{\relationwindowHeight}{1}
\pgfmathsetmacro{\gap}{1.8}
\pgfmathsetmacro{\windowHeight}{2.7}

\pgfmathsetmacro{\yAxis}{0}
\pgfmathsetmacro{\yPosA}{\yAxis  + 0.6}
\pgfmathsetmacro{\yPosB}{\yAxis  + 2*0.7 + 0.3}
\pgfmathsetmacro{\yPosC}{\yAxis  + 3*0.7 + 0.3}

 \draw (-0.7,\yAxis) [->, line width = 1pt]
    -- coordinate (x axis mid) (5,\yAxis);

\pgfmathsetmacro{\winTs}{-0.7}
\pgfmathsetmacro{\winTe}{3.7}
\draw [ultra thick, draw=none, pattern color=c1, pattern=north west lines, opacity=0.4]
(\winTs,\yAxis) -- (\winTs,\yAxis + \windowHeight - 0.2) -- (\winTe,\yAxis + \windowHeight - 0.2) -- (\winTe, \yAxis + 0)  -- cycle;

\draw [line width=1.2] (0, \yPosB) -- (4.5, \yPosB);

\draw [line width=3] (0.5,\yAxis+0.6) -- (1.7,\yAxis+0.6)
 node[pos=1,right=-1pt] {};
\draw [line width=3, color=yellow] (0.5,\yAxis+0.4) -- (1.7,\yAxis+0.4)
node[pos=1,right=-1pt] {};

\draw [line width=3] (2.6,\yAxis+0.6) -- (3.7,\yAxis+0.6)
 node[pos=1,right=-1pt] {};
\draw [line width=3, color=c4] (2.6,\yAxis+0.4) -- (3.7,\yAxis+0.4)
node[pos=1,right=-1pt] {};

\pgfmathsetmacro{\winTs}{3.7}
\draw [<-,line width=1,densely dashdotted,c1]
(\winTs, \yAxis) -- (\winTs,\yAxis +\windowHeight)
node[pos=1.1]
{\large {\color{c1} $\mathtt{windTs}$}};

\pgfmathsetmacro{\winTe}{4.5}
\draw [<-,line width=1,densely dashdotted,c3]
(\winTe, \yAxis) -- (\winTe,\yAxis +\windowHeight-0.45)
node[pos=1.1]
{\large {\color{c3} $\mathtt{windTe}$}};
 
\end{tikzpicture}}
\caption{Case 4}
\end{subfigure}
\begin{subfigure}[b]{0.48\linewidth} \centering
\scalebox{0.45} {
\begin{tikzpicture}   

\pgfmathsetmacro{\xAxisOne}{-0.2}
\pgfmathsetmacro{\xAxisTwo}{13.7}
\pgfmathsetmacro{\relationwindowHeight}{1}
\pgfmathsetmacro{\gap}{1.8}
\pgfmathsetmacro{\windowHeight}{2.7}

\pgfmathsetmacro{\yAxis}{0}
\pgfmathsetmacro{\yPosA}{\yAxis  + 0.6}
\pgfmathsetmacro{\yPosB}{\yAxis  + 2*0.7 + 0.3}
\pgfmathsetmacro{\yPosC}{\yAxis  + 3*0.7 + 0.3}

 \draw (-0.7,\yAxis) [->, line width = 1pt]
    -- coordinate (x axis mid) (5,\yAxis);
  
\pgfmathsetmacro{\winTs}{-0.7}
\pgfmathsetmacro{\winTe}{0}
\draw [ultra thick, draw=none, pattern color=c1, pattern=north west lines, opacity=0.4]
(\winTs,\yAxis) -- (\winTs,\yAxis + \windowHeight - 0.2) -- (\winTe,\yAxis + \windowHeight - 0.2) -- (\winTe, \yAxis + 0)  -- cycle;

\draw [line width=1.2] (0, \yPosB) -- (4.5, \yPosB);

\pgfmathsetmacro{\winTs}{\winTe}
\draw [<-,line width=1,densely dashdotted,c1]
(\winTs, \yAxis) -- (\winTs,\yAxis +\windowHeight)
node[pos=1.1]
{\large {\color{c1} $\mathtt{windTs}$}};

\pgfmathsetmacro{\winTs}{0}
\pgfmathsetmacro{\winTe}{4.5}


\draw [<-,line width=1,densely dashdotted,c3]
(\winTe, \yAxis) -- (\winTe,\yAxis +\windowHeight)
node[pos=1.1]
{\large {\color{c3} $\mathtt{windTe}$}};
 
\end{tikzpicture}}
\caption{Case 5}
\end{subfigure}
\caption{Cases for determining $\mathtt{windTe}$ in LAWA$_U$
Algorithm. Single line is used for the input tuple and pairs of lines for the windows.}
\label{fig:algoCases}
\end{figure}

%% file: 7a_fig_LAWAu.tex
\begin{figure}[h]\centering
\scalebox{0.55} {
\begin{tikzpicture}
\draw (0,0) [->, line width = 1pt]-- coordinate (x axis mid) (11.4,0);
  
\pgfmathsetmacro{\shift}{0.5}
\foreach \j in {1,...,8}{
\pgfmathsetmacro{\divRes}{int(\j/2)}
\pgfmathsetmacro{\modRes}{1-(\j-\divRes*2)}
\pgfmathsetmacro{\xPos}{\j-1+(\divRes-\modRes*\shift)}
\draw (\xPos , 1pt)  --  (\xPos ,-3pt)
node[anchor=north,font=\relsize{2}] {};
}

\pgfmathsetmacro{\shift}{0.5}
\foreach \j in {1,...,7}{
\pgfmathsetmacro{\divRes}{int(\j/2)}
\pgfmathsetmacro{\modRes}{1-(\j-\divRes*2)}
\pgfmathsetmacro{\xPos}{\j-1+(\divRes-\modRes*\shift)}
\draw (\xPos+0.75 , -3pt)
node[anchor=north,font=\relsize{1}] {\bf \j};
}

\pgfmathsetmacro{\xAxisOne}{-0.2}
\pgfmathsetmacro{\xAxisTwo}{12.3}

\pgfmathsetmacro{\windowHeight}{1.5}
\pgfmathsetmacro{\yAxis}{0.1}
\draw [line width=3] (4.5,\yAxis+0.6) -- (7.5,\yAxis+0.6)
 node[pos=1,right=-1pt] {};
\draw [line width=3, color=c5] (4.5,\yAxis+0.4) -- (7.5,\yAxis+0.4)
node[pos=1,right=-1pt] {};
\draw[snake=brace] (4.4,\yAxis+0.2) -- (4.4,\yAxis+0.8)
node[pos=0.5, left=2.5pt] {\large $x_1$};


\pgfmathsetmacro{\yAxis}{\yAxis + 0.7}
\draw [line width=3] (6,\yAxis+0.6) -- (10.5,\yAxis+0.6)
 node[pos=1,right=-1pt] {};
\draw [line width=3, color=c4] (6,\yAxis+0.4) -- (10.5,\yAxis+0.4)
node[pos=1,right=-1pt] {};
\draw[snake=brace] (5.9,\yAxis+0.2) -- (5.9,\yAxis+0.8)
node[pos=0.5, left=2.5pt] {\large $x_2$};


\pgfmathsetmacro{\yAxis}{\yAxis+1.2} 
\pgfmathsetmacro{\windowHeight}{0.9}
\draw [line width=1.5, ] (1.5, \yAxis) -- (10.5, \yAxis)
node[pos=.2,above=-1pt]{\large }; 

\pgfmathsetmacro{\yAxis}{0}
\pgfmathsetmacro{\windowHeight}{2.4}
\pgfmathsetmacro{\windTs}{1.5}
\pgfmathsetmacro{\windTe}{4.5}

\draw [<-,line width=1,densely dashdotted,color=c1]
(\windTs, \yAxis) -- (\windTs,\yAxis+\windowHeight)
node[pos=1.2]
{};
\draw [<-,line width=1,densely dashdotted,color=c1]
(\windTe, \yAxis) -- (\windTe,\yAxis +\windowHeight)
node[pos=1.2]
{};

\draw [ultra thick, draw=none, pattern color=c1, pattern=north west lines, opacity=0.4]
(\windTs,\yAxis) -- (\windTs,\yAxis + \windowHeight - 0.2) -- (\windTe,\yAxis + \windowHeight - 0.2) -- (\windTe, \yAxis + 0)  -- cycle;

\draw [thick,color=c1,decorate,
decoration={brace,amplitude=10pt,mirror},
xshift=-1pt,yshift=-1pt]
(\windTs,\yAxis-0.5) -- (\windTe,\yAxis-0.5)
node [c1,pos=0.85,yshift=-0.7cm] 
{\large $\mathtt{out}$ = $(\bsq{Ann, ZAK}, \mathtt{null}, [2,4), a_1, \mathtt{null})$};

\pgfmathsetmacro{\yAxis}{4.3}
\draw (0,\yAxis) [->, line width = 1pt]-- 
coordinate (x axis mid) (11.4,\yAxis);
  
\pgfmathsetmacro{\shift}{0.5}
\foreach \j in {1,...,8}{
\pgfmathsetmacro{\divRes}{int(\j/2)}
\pgfmathsetmacro{\modRes}{1-(\j-\divRes*2)}
\pgfmathsetmacro{\xPos}{\j-1+(\divRes-\modRes*\shift)}
\draw (\xPos , \yAxis +0.1)  --  (\xPos , \yAxis-0.1)
node[anchor=north,font=\relsize{2}] {};
}

\pgfmathsetmacro{\shift}{0.5}
\foreach \j in {1,...,7}{
\pgfmathsetmacro{\divRes}{int(\j/2)}
\pgfmathsetmacro{\modRes}{1-(\j-\divRes*2)}
\pgfmathsetmacro{\xPos}{\j-1+(\divRes-\modRes*\shift)}
\draw (\xPos+0.75 , \yAxis-0.1)
node[anchor=north,font=\relsize{1}] {\bf \j};
}   

\pgfmathsetmacro{\xAxisOne}{-0.2}
\pgfmathsetmacro{\xAxisTwo}{12.3}
 
\pgfmathsetmacro{\windowHeight}{1.5}
\pgfmathsetmacro{\yAxis}{\yAxis + 0.1}
\draw [line width=3] (4.5,\yAxis+0.6) -- (7.5,\yAxis+0.6)
 node[pos=1,right=-1pt] {};
\draw [line width=3, color=c5] (4.5,\yAxis+0.4) -- (7.5,\yAxis+0.4)
node[pos=1,right=-1pt] {};
\draw[snake=brace] (4.4,\yAxis+0.2) -- (4.4,\yAxis+0.8)
node[pos=0.5, left=2.5pt] {\large $x_1$};


\pgfmathsetmacro{\yAxis}{\yAxis + 0.7}
\draw [line width=3] (6,\yAxis+0.6) -- (10.5,\yAxis+0.6)
 node[pos=1,right=-1pt] {};
\draw [line width=3, color=c4] (6,\yAxis+0.4) -- (10.5,\yAxis+0.4)
node[pos=1,right=-1pt] {};
\draw[snake=brace] (5.9,\yAxis+0.2) -- (5.9,\yAxis+0.8)
node[pos=0.5, left=2.5pt] {\large $x_2$};

\pgfmathsetmacro{\yAxis}{\yAxis+1.2} 
\pgfmathsetmacro{\windowHeight}{0.9}
\draw [line width=1.5] (1.5, \yAxis) -- (10.5, \yAxis)
node[pos=.2,above=-1pt]{\large }; 

\pgfmathsetmacro{\yAxis}{4.3}
\pgfmathsetmacro{\windowHeight}{2.4}
\pgfmathsetmacro{\windTs}{4.5}
\pgfmathsetmacro{\windTe}{7.5}

\draw [<-,line width=1,densely dashdotted,red]
(\windTs, \yAxis) -- (\windTs,\yAxis+\windowHeight)
node[pos=1.2]
{};
\draw [<-,line width=1,densely dashdotted,red]
(\windTe, \yAxis) -- (\windTe,\yAxis +\windowHeight)
node[pos=1.2]
{};

\draw [ultra thick, draw=none, pattern color=red, pattern=north west lines, opacity=0.4]
(\windTs,\yAxis) -- (\windTs,\yAxis + \windowHeight - 0.2) -- (\windTe,\yAxis + \windowHeight - 0.2) -- (\windTe, \yAxis + 0)  -- cycle;

\draw [thick,red,decorate,
decoration={brace,amplitude=10pt,mirror},
xshift=-1pt,yshift=-1pt]
(\windTs,\yAxis-0.5) -- (\windTe,\yAxis-0.5)
node [red,pos=0.15,yshift=-0.7cm] 
{\large $\mathtt{out}$ =  $(\bsq{Ann, ZAK}, \bsq{hotel_1, ZAK}, [4,6), a_1, b_3)$};

\end{tikzpicture}}
\caption{LAWA$_{U}$ on the group with $\mathtt{F_L} = \bsq{Ann, ZAK}$ and $\mathtt{\lambda_L}=a_1$.}
\label{fig:algoUOrun}
\end{figure}
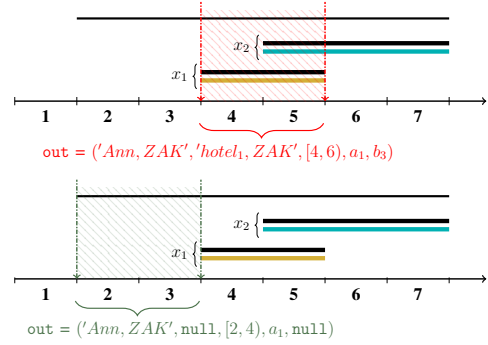

%% file: 7a_fig_LAWAn.tex
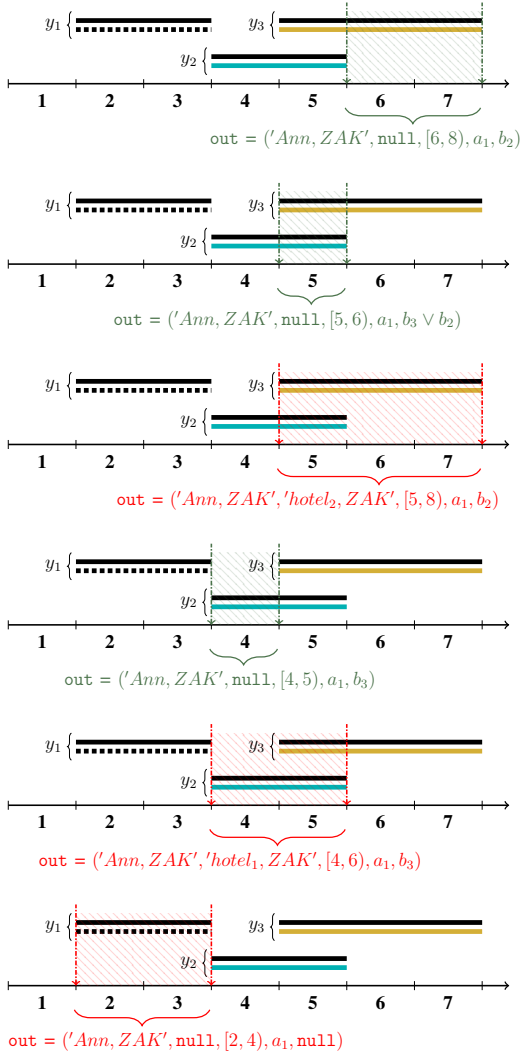
\begin{figure}[!ht]\centering
\scalebox{0.6} {
\begin{tikzpicture}
\pgfmathsetmacro{\xAxisOne}{-0.2}
\pgfmathsetmacro{\xAxisTwo}{12.3}
\pgfmathsetmacro{\windowHeight}{1.8}

\pgfmathsetmacro{\yAxisWinOne}{0}
\pgfmathsetmacro{\yAxisWinTwo}{4}
\pgfmathsetmacro{\yAxisWinThree}{8}
\pgfmathsetmacro{\yAxisWinFour}{12}
\pgfmathsetmacro{\yAxisWinFive}{16}
\pgfmathsetmacro{\yAxisWinSix}{20}
\pgfmathsetmacro{\yAxisWinSeven}{24}

\pgfmathsetmacro{\yAxis}{\yAxisWinOne}
\draw (0,\yAxis) [->, line width = 1pt]-- 
coordinate (x axis mid) (11.1,\yAxis);
  
\pgfmathsetmacro{\shift}{0.5}
\foreach \j in {1,...,8}{
\pgfmathsetmacro{\divRes}{int(\j/2)}
\pgfmathsetmacro{\modRes}{1-(\j-\divRes*2)}
\pgfmathsetmacro{\xPos}{\j-1+(\divRes-\modRes*\shift)}
\draw (\xPos , \yAxis +0.1)  --  (\xPos , \yAxis-0.1)
node[anchor=north,font=\relsize{2}] {};
}

\pgfmathsetmacro{\shift}{0.5}
\foreach \j in {1,...,7}{
\pgfmathsetmacro{\divRes}{int(\j/2)}
\pgfmathsetmacro{\modRes}{1-(\j-\divRes*2)}
\pgfmathsetmacro{\xPos}{\j-1+(\divRes-\modRes*\shift)}
\draw (\xPos+0.75 , \yAxis-0.1)
node[anchor=north,font=\relsize{1}] {\bf \j};
}

\draw [line width=3] (1.5,\yAxis+1.4) -- (4.5,\yAxis+1.4)
 node[pos=1,right=-1pt] {};
\draw [line width=3,dotted] (1.5,\yAxis+1.2) -- (4.5,\yAxis+1.2)
node[pos=1,right=-1pt] {}; 
\draw[snake=brace] (1.4,\yAxis+1) -- (1.4,\yAxis+1.6)node[pos=0.5, left=2.5pt] {\large $y_1$};

\draw [line width=3] (6,\yAxis+1.4) -- (10.5,\yAxis+1.4)
 node[pos=1,right=-1pt] {};
\draw [line width=3, color=c5] (6,\yAxis+1.2) -- (10.5,\yAxis+1.2)
node[pos=1,right=-1pt] {}; 
\draw[snake=brace] (5.9,\yAxis+1) -- (5.9,\yAxis+1.6)node[pos=0.5, left=2.5pt] {\large $y_3$};

\draw [line width=3] (4.5,\yAxis+0.6) -- (7.5,\yAxis+0.6)
 node[pos=1,right=-1pt] {};
\draw [line width=3, color=c4] (4.5,\yAxis+0.4) -- (7.5,\yAxis+0.4)
node[pos=1,right=-1pt] {}; 
\draw[snake=brace] (4.4,\yAxis+0.2) -- (4.4,\yAxis+0.8)node[pos=0.5, left=2.5pt] {\large $y_2$};

\pgfmathsetmacro{\yAxis}{\yAxisWinOne}
\pgfmathsetmacro{\windTs}{1.5}
\pgfmathsetmacro{\windTe}{4.5}

\draw [<-,line width=1,densely dashdotted,red]
(\windTs, \yAxis) -- (\windTs,\yAxis+\windowHeight)
node[pos=1.2] {};
\draw [<-,line width=1,densely dashdotted,red]
(\windTe, \yAxis) -- (\windTe,\yAxis +\windowHeight)
node[pos=1.2] {};

\draw [ultra thick, draw=none, pattern color=red, pattern=north west lines, opacity=0.4]
(\windTs,\yAxis) -- (\windTs,\yAxis + \windowHeight - 0.2) -- (\windTe,\yAxis + \windowHeight - 0.2) -- (\windTe, \yAxis + 0)  -- cycle;

\draw [thick,red,decorate,
decoration={brace,amplitude=10pt,mirror},
xshift=-1pt,yshift=-1pt]
(\windTs,\yAxis-0.5) -- (\windTe,\yAxis-0.5)
node [red,pos=0.75,yshift=-0.7cm] 
{\large $\mathtt{out}$ = $(\bsq{Ann, ZAK}, \mathtt{null}, [2,4), a_1, \mathtt{null})$};

\pgfmathsetmacro{\yAxis}{\yAxisWinTwo}
\draw (0,\yAxis) [->, line width = 1pt]-- 
coordinate (x axis mid) (11.1,\yAxis);
  
\pgfmathsetmacro{\shift}{0.5}
\foreach \j in {1,...,8}{
\pgfmathsetmacro{\divRes}{int(\j/2)}
\pgfmathsetmacro{\modRes}{1-(\j-\divRes*2)}
\pgfmathsetmacro{\xPos}{\j-1+(\divRes-\modRes*\shift)}
\draw (\xPos , \yAxis +0.1)  --  (\xPos , \yAxis-0.1)
node[anchor=north,font=\relsize{2}] {};
}

\pgfmathsetmacro{\shift}{0.5}
\foreach \j in {1,...,7}{
\pgfmathsetmacro{\divRes}{int(\j/2)}
\pgfmathsetmacro{\modRes}{1-(\j-\divRes*2)}
\pgfmathsetmacro{\xPos}{\j-1+(\divRes-\modRes*\shift)}
\draw (\xPos+0.75 , \yAxis-0.1)
node[anchor=north,font=\relsize{1}] {\bf \j};
}

\draw [line width=3] (1.5,\yAxis+1.4) -- (4.5,\yAxis+1.4)
 node[pos=1,right=-1pt] {};
\draw [line width=3,dotted] (1.5,\yAxis+1.2) -- (4.5,\yAxis+1.2)
node[pos=1,right=-1pt] {}; 
\draw[snake=brace] (1.4,\yAxis+1) -- (1.4,\yAxis+1.6)node[pos=0.5, left=2.5pt] {\large $y_1$};

\draw [line width=3] (6,\yAxis+1.4) -- (10.5,\yAxis+1.4)
 node[pos=1,right=-1pt] {};
\draw [line width=3, color=c5] (6,\yAxis+1.2) -- (10.5,\yAxis+1.2)
node[pos=1,right=-1pt] {}; 
\draw[snake=brace] (5.9,\yAxis+1) -- (5.9,\yAxis+1.6)node[pos=0.5, left=2.5pt] {\large $y_3$};

\draw [line width=3] (4.5,\yAxis+0.6) -- (7.5,\yAxis+0.6)
 node[pos=1,right=-1pt] {};
\draw [line width=3, color=c4] (4.5,\yAxis+0.4) -- (7.5,\yAxis+0.4)
node[pos=1,right=-1pt] {}; 
\draw[snake=brace] (4.4,\yAxis+0.2) -- (4.4,\yAxis+0.8)node[pos=0.5, left=2.5pt] {\large $y_2$};

\pgfmathsetmacro{\yAxis}{\yAxisWinTwo}
\pgfmathsetmacro{\windTs}{4.5}
\pgfmathsetmacro{\windTe}{7.5}

\draw [<-,line width=1,densely dashdotted,red]
(\windTs, \yAxis) -- (\windTs,\yAxis+\windowHeight)
node[pos=1.2] {};
\draw [<-,line width=1,densely dashdotted,red]
(\windTe, \yAxis) -- (\windTe,\yAxis +\windowHeight)
node[pos=1.2] {};

\draw [ultra thick, draw=none, pattern color=red, pattern=north west lines, opacity=0.4]
(\windTs,\yAxis) -- (\windTs,\yAxis + \windowHeight - 0.2) -- (\windTe,\yAxis + \windowHeight - 0.2) -- (\windTe, \yAxis + 0)  -- cycle;

\draw [thick,red,decorate,
decoration={brace,amplitude=10pt,mirror},
xshift=-1pt,yshift=-1pt]
(\windTs,\yAxis-0.5) -- (\windTe,\yAxis-0.5)
node [red,pos=0.15,yshift=-0.7cm] 
{\large $\mathtt{out}$ = $(\bsq{Ann, ZAK}, \bsq{hotel_1, ZAK}, [4,6), a_1, b_3)$};

\pgfmathsetmacro{\yAxis}{\yAxisWinThree}
\draw (0,\yAxis) [->, line width = 1pt]-- 
coordinate (x axis mid) (11.1,\yAxis);
  
\pgfmathsetmacro{\shift}{0.5}
\foreach \j in {1,...,8}{
\pgfmathsetmacro{\divRes}{int(\j/2)}
\pgfmathsetmacro{\modRes}{1-(\j-\divRes*2)}
\pgfmathsetmacro{\xPos}{\j-1+(\divRes-\modRes*\shift)}
\draw (\xPos , \yAxis +0.1)  --  (\xPos , \yAxis-0.1)
node[anchor=north,font=\relsize{2}] {};
}

\pgfmathsetmacro{\shift}{0.5}
\foreach \j in {1,...,7}{
\pgfmathsetmacro{\divRes}{int(\j/2)}
\pgfmathsetmacro{\modRes}{1-(\j-\divRes*2)}
\pgfmathsetmacro{\xPos}{\j-1+(\divRes-\modRes*\shift)}
\draw (\xPos+0.75 , \yAxis-0.1)
node[anchor=north,font=\relsize{1}] {\bf \j};
}

\draw [line width=3] (1.5,\yAxis+1.4) -- (4.5,\yAxis+1.4)
 node[pos=1,right=-1pt] {};
\draw [line width=3,dotted] (1.5,\yAxis+1.2) -- (4.5,\yAxis+1.2)
node[pos=1,right=-1pt] {}; 
\draw[snake=brace] (1.4,\yAxis+1) -- (1.4,\yAxis+1.6)node[pos=0.5, left=2.5pt] {\large $y_1$};

\draw [line width=3] (6,\yAxis+1.4) -- (10.5,\yAxis+1.4)
 node[pos=1,right=-1pt] {};
\draw [line width=3, color=c5] (6,\yAxis+1.2) -- (10.5,\yAxis+1.2)
node[pos=1,right=-1pt] {}; 
\draw[snake=brace] (5.9,\yAxis+1) -- (5.9,\yAxis+1.6)node[pos=0.5, left=2.5pt] {\large $y_3$};

\draw [line width=3] (4.5,\yAxis+0.6) -- (7.5,\yAxis+0.6)
 node[pos=1,right=-1pt] {};
\draw [line width=3, color=c4] (4.5,\yAxis+0.4) -- (7.5,\yAxis+0.4)
node[pos=1,right=-1pt] {}; 
\draw[snake=brace] (4.4,\yAxis+0.2) -- (4.4,\yAxis+0.8)node[pos=0.5, left=2.5pt] {\large $y_2$};

\pgfmathsetmacro{\yAxis}{\yAxisWinThree}
\pgfmathsetmacro{\windTs}{4.5}
\pgfmathsetmacro{\windTe}{6}

\draw [<-,line width=1,densely dashdotted,c1]
(\windTs, \yAxis) -- (\windTs,\yAxis+\windowHeight)
node[pos=1.2] {};
\draw [<-,line width=1,densely dashdotted,c1]
(\windTe, \yAxis) -- (\windTe,\yAxis +\windowHeight)
node[pos=1.2] {};

\draw [ultra thick, draw=none, pattern color=c1, pattern=north west lines, opacity=0.4]
(\windTs,\yAxis) -- (\windTs,\yAxis + \windowHeight - 0.2) -- (\windTe,\yAxis + \windowHeight - 0.2) -- (\windTe, \yAxis + 0)  -- cycle;

\draw [thick,c1,decorate,
decoration={brace,amplitude=10pt,mirror},
xshift=-1pt,yshift=-1pt]
(\windTs,\yAxis-0.5) -- (\windTe,\yAxis-0.5)
node [c1,pos=0.15,yshift=-0.7cm] 
{\large $\mathtt{out}$ = $(\bsq{Ann, ZAK}, \mathtt{null}, [4,5), a_1, b_3)$};

\pgfmathsetmacro{\yAxis}{\yAxisWinFour}
\draw (0,\yAxis) [->, line width = 1pt]-- coordinate (x axis mid) (11.1,\yAxis);
  
\pgfmathsetmacro{\shift}{0.5}
\foreach \j in {1,...,8}{
\pgfmathsetmacro{\divRes}{int(\j/2)}
\pgfmathsetmacro{\modRes}{1-(\j-\divRes*2)}
\pgfmathsetmacro{\xPos}{\j-1+(\divRes-\modRes*\shift)}
\draw (\xPos , \yAxis+0.1)  --  (\xPos ,\yAxis-0.1)
node[anchor=north,font=\relsize{2}] {};
}

\pgfmathsetmacro{\shift}{0.5}
\foreach \j in {1,...,7}{
\pgfmathsetmacro{\divRes}{int(\j/2)}
\pgfmathsetmacro{\modRes}{1-(\j-\divRes*2)}
\pgfmathsetmacro{\xPos}{\j-1+(\divRes-\modRes*\shift)}
\draw (\xPos+0.75 , \yAxis-0.1)
node[anchor=north,font=\relsize{1}] {\bf \j};
}

\pgfmathsetmacro{\xAxisOne}{-0.2}
\pgfmathsetmacro{\xAxisTwo}{12.3}
 
\draw [line width=3] (1.5,\yAxis+1.4) -- (4.5,\yAxis+1.4)
 node[pos=1,right=-1pt] {};
\draw [line width=3,dotted] (1.5,\yAxis+1.2) -- (4.5,\yAxis+1.2)
node[pos=1,right=-1pt] {}; 
\draw[snake=brace] (1.4,\yAxis+1) -- (1.4,\yAxis+1.6)node[pos=0.5, left=2.5pt] {\large $y_1$};

\draw [line width=3] (6,\yAxis+1.4) -- (10.5,\yAxis+1.4)
 node[pos=1,right=-1pt] {};
\draw [line width=3, color=c5] (6,\yAxis+1.2) -- (10.5,\yAxis+1.2)
node[pos=1,right=-1pt] {}; 
\draw[snake=brace] (5.9,\yAxis+1) -- (5.9,\yAxis+1.6)node[pos=0.5, left=2.5pt] {\large $y_3$};

\draw [line width=3] (4.5,\yAxis+0.6) -- (7.5,\yAxis+0.6)
 node[pos=1,right=-1pt] {};
\draw [line width=3, color=c4] (4.5,\yAxis+0.4) -- (7.5,\yAxis+0.4)
node[pos=1,right=-1pt] {}; 
\draw[snake=brace] (4.4,\yAxis+0.2) -- (4.4,\yAxis+0.8)node[pos=0.5, left=2.5pt] {\large $y_2$};

\pgfmathsetmacro{\yAxis}{\yAxisWinFour}
\pgfmathsetmacro{\windTs}{6}
\pgfmathsetmacro{\windTe}{10.5}

\draw [<-,line width=1,densely dashdotted,red]
(\windTs, \yAxis) -- (\windTs,\yAxis+\windowHeight)
node[pos=1.2] {};
\draw [<-,line width=1,densely dashdotted,red]
(\windTe, \yAxis) -- (\windTe,\yAxis +\windowHeight)
node[pos=1.2] {};

\draw [ultra thick, draw=none, pattern color=red, pattern=north west lines, opacity=0.4]
(\windTs,\yAxis) -- (\windTs,\yAxis + \windowHeight - 0.2) -- (\windTe,\yAxis + \windowHeight - 0.2) -- (\windTe, \yAxis + 0)  -- cycle;

\draw [thick,red,decorate,
decoration={brace,amplitude=10pt,mirror},
xshift=-1pt,yshift=-1pt]
(\windTs,\yAxis-0.5) -- (\windTe,\yAxis-0.5)
node [red,pos=0.15,yshift=-0.7cm] 
{\large $\mathtt{out}$ = $(\bsq{Ann, ZAK}, \bsq{hotel_2, ZAK}, [5,8), a_1, b_2)$};

\pgfmathsetmacro{\yAxis}{\yAxisWinFive}
\draw (0,\yAxis) [->, line width = 1pt]-- coordinate (x axis mid) (11.1,\yAxis);
  
\pgfmathsetmacro{\shift}{0.5}
\foreach \j in {1,...,8}{
\pgfmathsetmacro{\divRes}{int(\j/2)}
\pgfmathsetmacro{\modRes}{1-(\j-\divRes*2)}
\pgfmathsetmacro{\xPos}{\j-1+(\divRes-\modRes*\shift)}
\draw (\xPos , \yAxis+0.1)  --  (\xPos ,\yAxis-0.1)
node[anchor=north,font=\relsize{2}] {};
}

\pgfmathsetmacro{\shift}{0.5}
\foreach \j in {1,...,7}{
\pgfmathsetmacro{\divRes}{int(\j/2)}
\pgfmathsetmacro{\modRes}{1-(\j-\divRes*2)}
\pgfmathsetmacro{\xPos}{\j-1+(\divRes-\modRes*\shift)}
\draw (\xPos+0.75 , \yAxis-0.1)
node[anchor=north,font=\relsize{1}] {\bf \j};
}

\pgfmathsetmacro{\xAxisOne}{-0.2}
\pgfmathsetmacro{\xAxisTwo}{12.3}
 
\draw [line width=3] (1.5,\yAxis+1.4) -- (4.5,\yAxis+1.4)
 node[pos=1,right=-1pt] {};
\draw [line width=3,dotted] (1.5,\yAxis+1.2) -- (4.5,\yAxis+1.2)
node[pos=1,right=-1pt] {}; 
\draw[snake=brace] (1.4,\yAxis+1) -- (1.4,\yAxis+1.6)node[pos=0.5, left=2.5pt] {\large $y_1$};

\draw [line width=3] (6,\yAxis+1.4) -- (10.5,\yAxis+1.4)
 node[pos=1,right=-1pt] {};
\draw [line width=3, color=c5] (6,\yAxis+1.2) -- (10.5,\yAxis+1.2)
node[pos=1,right=-1pt] {}; 
\draw[snake=brace] (5.9,\yAxis+1) -- (5.9,\yAxis+1.6)node[pos=0.5, left=2.5pt] {\large $y_3$};

\draw [line width=3] (4.5,\yAxis+0.6) -- (7.5,\yAxis+0.6)
 node[pos=1,right=-1pt] {};
\draw [line width=3, color=c4] (4.5,\yAxis+0.4) -- (7.5,\yAxis+0.4)
node[pos=1,right=-1pt] {}; 
\draw[snake=brace] (4.4,\yAxis+0.2) -- (4.4,\yAxis+0.8)node[pos=0.5, left=2.5pt] {\large $y_2$};

\pgfmathsetmacro{\yAxis}{\yAxisWinFive}
\pgfmathsetmacro{\windTs}{6}
\pgfmathsetmacro{\windTe}{7.5}

\draw [<-,line width=1,densely dashdotted,c1]
(\windTs, \yAxis) -- (\windTs,\yAxis+\windowHeight)
node[pos=1.2] {};
\draw [<-,line width=1,densely dashdotted,c1]
(\windTe, \yAxis) -- (\windTe,\yAxis +\windowHeight)
node[pos=1.2] {};

\draw [ultra thick, draw=none, pattern color=c1, pattern=north west lines, opacity=0.4]
(\windTs,\yAxis) -- (\windTs,\yAxis + \windowHeight - 0.2) -- (\windTe,\yAxis + \windowHeight - 0.2) -- (\windTe, \yAxis + 0)  -- cycle;

\draw [thick,c1,decorate,
decoration={brace,amplitude=10pt,mirror},
xshift=-1pt,yshift=-1pt]
(\windTs,\yAxis-0.5) -- (\windTe,\yAxis-0.5)
node [c1,pos=0.15,yshift=-0.7cm] 
{\large $\mathtt{out}$ = $(\bsq{Ann, ZAK}, \mathtt{null}, [5,6), a_1, b_3 \lor b_2)$};

\pgfmathsetmacro{\yAxis}{\yAxisWinSix}
\draw (0,\yAxis) [->, line width = 1pt]-- 
coordinate (x axis mid) (11.1,\yAxis);
  
\pgfmathsetmacro{\shift}{0.5}
\foreach \j in {1,...,8}{
\pgfmathsetmacro{\divRes}{int(\j/2)}
\pgfmathsetmacro{\modRes}{1-(\j-\divRes*2)}
\pgfmathsetmacro{\xPos}{\j-1+(\divRes-\modRes*\shift)}
\draw (\xPos , \yAxis +0.1)  --  (\xPos , \yAxis-0.1)
node[anchor=north,font=\relsize{2}] {};
}

\pgfmathsetmacro{\shift}{0.5}
\foreach \j in {1,...,7}{
\pgfmathsetmacro{\divRes}{int(\j/2)}
\pgfmathsetmacro{\modRes}{1-(\j-\divRes*2)}
\pgfmathsetmacro{\xPos}{\j-1+(\divRes-\modRes*\shift)}
\draw (\xPos+0.75 , \yAxis-0.1)
node[anchor=north,font=\relsize{1}] {\bf \j};
}

\draw [line width=3] (1.5,\yAxis+1.4) -- (4.5,\yAxis+1.4)
 node[pos=1,right=-1pt] {};
\draw [line width=3,dotted] (1.5,\yAxis+1.2) -- (4.5,\yAxis+1.2)
node[pos=1,right=-1pt] {}; 
\draw[snake=brace] (1.4,\yAxis+1) -- (1.4,\yAxis+1.6)node[pos=0.5, left=2.5pt] {\large $y_1$};

\draw [line width=3] (6,\yAxis+1.4) -- (10.5,\yAxis+1.4)
 node[pos=1,right=-1pt] {};
\draw [line width=3, color=c5] (6,\yAxis+1.2) -- (10.5,\yAxis+1.2)
node[pos=1,right=-1pt] {}; 
\draw[snake=brace] (5.9,\yAxis+1) -- (5.9,\yAxis+1.6)node[pos=0.5, left=2.5pt] {\large $y_3$};

\draw [line width=3] (4.5,\yAxis+0.6) -- (7.5,\yAxis+0.6)
 node[pos=1,right=-1pt] {};
\draw [line width=3, color=c4] (4.5,\yAxis+0.4) -- (7.5,\yAxis+0.4)
node[pos=1,right=-1pt] {}; 
\draw[snake=brace] (4.4,\yAxis+0.2) -- (4.4,\yAxis+0.8)node[pos=0.5, left=2.5pt] {\large $y_2$};

\pgfmathsetmacro{\yAxis}{\yAxisWinSix}
\pgfmathsetmacro{\windTs}{7.5}
\pgfmathsetmacro{\windTe}{10.5}

\draw [<-,line width=1,densely dashdotted,c1]
(\windTs, \yAxis) -- (\windTs,\yAxis+\windowHeight)
node[pos=1.2] {};
\draw [<-,line width=1,densely dashdotted,c1]
(\windTe, \yAxis) -- (\windTe,\yAxis +\windowHeight)
node[pos=1.2] {};

\draw [ultra thick, draw=none, pattern color=c1, pattern=north west lines, opacity=0.4]
(\windTs,\yAxis) -- (\windTs,\yAxis + \windowHeight - 0.2) -- (\windTe,\yAxis + \windowHeight - 0.2) -- (\windTe, \yAxis + 0)  -- cycle;

\draw [thick,c1,decorate,
decoration={brace,amplitude=10pt,mirror},
xshift=-1pt,yshift=-1pt]
(\windTs,\yAxis-0.5) -- (\windTe,\yAxis-0.5)
node [c1,pos=0.15,yshift=-0.7cm] 
{\large $\mathtt{out}$ = $(\bsq{Ann, ZAK}, \mathtt{null}, [6,8), a_1, b_2)$};

\end{tikzpicture}}
\caption{Execution of LAWA$_\kat{N}$ on the result of LAWA$_\kat{U}$}
\label{fig:algoNrun}
\end{figure}

%% file: 8_experimental_evaluation.tex
In this section, we evaluate our algorithms using two real-world
datasets which vary on (i) the number of facts in the input
relations and (ii) the percentage of tuples whose intervals
overlap. We compare our approach for TP joins with negation
(NJ) to Temporal Alignment (TA), i.e., the only related
approach that can be used for the computation of TP
outer joins and TP anti join. The experiments show that our
approach outperforms TA and it is the only scalable solution
for TP joins with negation on input relations of more than
200K tuples. {\em NJ} is also robust with predictable
performance with respect to the aforementioned 
characteristics of the datasets.

\subsection{Experimental Setup}

All of the following experiments were deployed on a 2xIntel(R)
Xeon(R) CPU E5-24400 @2.40GHz machine with 64GB main memory,
running CentOS 6.7. Our algorithms have been implemented in the
kernel of PostgreSQL in C, and all experiments were performed in
main-memory. No indexes were used. In all PostgreSQL
implementations, the maximum memory for sorting as well as for
shared buffers were set to 10GB.


We have implemented {\em NJ} in PostgreSQL 9.4.3 by modifying the
parser, executor and optimizer. The only approach our
implementation can be compared against is \textbf{Temporal
Alignment (TA)~\cite{DignosTODS16}}. \textbf{Temporal Alignment}
is an approach developed for the computation of temporal
operations using sequenced semantics and is implemented in the
kernel of PostgreSQL as well. It consists of a set of reduction
rules based on \emph{Normalize} ($\mathcal{N}$) and \emph{Align}
($\Phi$), two operators responsible for the interval adjustment of
the input relations. Due to the existence of probabilities, the
results of TP joins with negation differ and thus, for our
experiments, we introduced reduction rules that are consistent
with the TP semantics while properly exploiting $\mathcal{N}$ and
$\Phi$. For a fair comparison, we migrated the authors'
implementation to PostgreSQL 9.4.3.


\tikzstyle{bag} = [text width=5em, text centered]
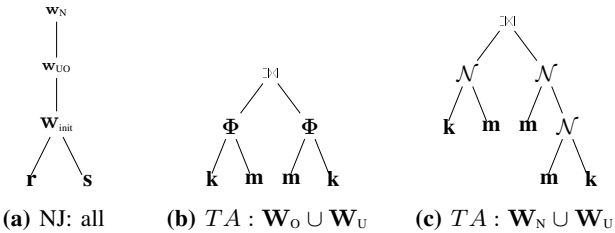
\begin{figure}[!ht]
\fontsize{8pt}{11pt}\selectfont
\setlength{\tabcolsep}{2pt}
\centering
\begin{subfigure} [b]{0.25\linewidth}
\centering
\scalebox{0.75}{
\begin{tikzpicture}[sloped]
\tikzstyle{level 1}=[level distance=1cm, sibling distance=1.5cm]
\tikzstyle{level 2}=[level distance=1cm, sibling distance=1cm]
\node[bag] {{$\mathbf{w_\kat{N}}$}}
 child  {
   node[bag] {{$\mathbf{w_\kat{UO}}$}}
     child {
       node[bag] {\large{$\mathbf{w_\kat{init}}$}}
       child {node[bag, label=center:{\textbf{ \large r}}] {}}
       child {node[bag, label=center:{\textbf{ \large s}}] {}}
        }
    } ;
\end{tikzpicture}}
\caption{NJ: all}
\label{fig:qTreeLAWA}
\end{subfigure}
\begin{subfigure} [b]{0.37\linewidth}
\centering
\scalebox{0.7}{
\begin{tikzpicture}[sloped]
\tikzstyle{level 1}=[level distance=1cm, sibling distance=1.5cm]
\tikzstyle{level 2}=[level distance=1cm, sibling distance=0.8cm]
\node[bag] {\large{$\mathbf{\LJoin}$}}                     
child  {                                                   
    node[bag] {\large{$\mathbf{\Phi}$}}                    
    child {node[bag, label=center:{\textbf{ \large k}}] {}}
    child {node[bag, label=center:{\textbf{ \large m}}] {}}
}                                                          
child  {                                                   
    node[bag] {\large{$\mathbf{\Phi}$}}                    
    child {node[bag, label=center:{\textbf{ \large m}}] {}}
    child {node[bag, label=center:{\textbf{ \large k}}] {}}
};                                                         

%
\end{tikzpicture}}
\caption{$TA: {\bf W_\kat{O}} \cup {\bf W_\kat{U}}$}
\label{fig:qTreeTAuo}
\end{subfigure}
\begin{subfigure} [b]{0.35\linewidth}
\centering
\scalebox{0.7}{
\begin{tikzpicture}[sloped]
\tikzstyle{level 1}=[level distance=1cm, sibling distance=1.5cm]
\tikzstyle{level 2}=[level distance=1cm, sibling distance=0.8cm]

\node[bag] {\large{$\mathbf{\LJoin}$}}
child  {
    node[bag] {\large{$\mathbf{\mathcal{N}}$}}
    child {node[bag, label=center:{\textbf{ \large k}}] {}}
    child {node[bag, label=center:{\textbf{ \large m}}] {}}
} 
child  {
    node[bag] {\large{$\mathbf{\mathcal{N}}$}}
	child {node[bag, label=center:{\textbf{ \large m}}] {}}
	child {
        node[bag] {\large{$\mathbf{\mathcal{N}}$}}
	    child {node[bag, label=center:{\textbf{ \large m}}] {}}
    	child {node[bag, label=center:{\textbf{ \large k}}] {}}
    }
};  
\end{tikzpicture}}
\caption{$TA: {\bf W_\kat{N}} \cup {\bf W_\kat{U}}$}
\label{fig:qTreeTAn}
\end{subfigure}
\caption{Query Trees}
\label{fig:qTrees}
\end{figure}

In Fig.~\ref{fig:qTrees}, we illustrate the query plans used by
NJ and TA for the computation of windows. In
Fig.~\ref{fig:qTreeLAWA}, the nodes $\mathtt{w_{init}}$, 
$\mathtt{w_{uo}}$ in the tree correspond to sets of windows as
described in Algorithm~\ref{algo:negatingJoins}. The node
$\mathtt{w_N}$ corresponds to the set of negating windows produced
by the calls of LAWA$_N$. In Fig.~\ref{fig:qTreeTAuo} and
\ref{fig:qTreeTAn}, we illustrate the two query subtrees in TA for
the computation of all output tuples. The operators $\mathcal{N}$
and $\Phi$ in TA replicate the tuples of the left relation and
assign new intervals based on the right relation. Since the facts
and lineages of the input tuples still need to be combined,
additional joins are performed. $\Phi(\mathbf{k},\mathbf{m})$ is
associated with overlapping windows (Fig.~\ref{fig:qTreeTAuo})
since the subintervals it produces correspond to the overlap of a
tuple in ${\bf k}$ with a tuple in ${\bf m}$.
$\mathcal{N}(\mathbf{k}, \mathbf{m})$ is appropriate for negating
windows since it includes intervals that correspond to the overlap
of a tuple in ${\bf k}$ with a group of tuples in ${\bf m}$. Both
$\Phi({\bf k},{\bf m})$ and $\mathcal{N} (\mathbf{k},\mathbf{m})$
include intervals where a tuple $k$ in $\mathbf{k}$ matches no
tuple in ${\bf m}$, leading to the unmatched windows being
computed twice. In Fig.~\ref{fig:qTreeTAn}, the tuples of the
right relation $\mathbf{m}$ are adjusted both using relation
$\mathbf{k}$ and itself because, over an interval, we compute the
tuples of $\mathbf{m}$ that are valid and are combined with a
tuple of $\mathbf{k}$. Given that $\mathcal{N}$ only uses one
input relation as reference, we need to further adjust
$\mathbf{m}$ based on the result of
$\mathcal{N} (\mathbf{k},\mathbf{m})$.

The $\LJoin_{\theta \land \theta_{o}}$, $\mathcal{N}$ and $\Phi$
nodes are all based on a conventional left-outer join with a
condition for the interval overlap of the matching tuples. 
PostgreSQL's optimizer determines whether such a join is executed
as a nested loop, a merge join or a hash join depending on the
$\theta$ codition of the TP join to be computed.
$\LJoin_{\theta \land \theta_{o}}$ is computed using a nested
loop only when the $\theta$ condition used has low selectivity,
i.e., when a high percentage of pairs of input tuples satisfy the
condition. On the contrary, this varies for $\mathcal{N}$ and
$\Phi$, based on whether a TP join or a set of windows is 
computed.



\subsection{Real-World Datasets}

The Webkit dataset\footnote{\scriptsize{The WebKit Open Source
Project: \url{http://www.webkit.org} (2012)}} \cite{DignosBG14,
PlatovICDE16, CafagnaB17} records the history of 484K files of 
the SVN repository of the Webkit project over a period of 11
years at a granularity of milliseconds. Each tuple has schema
(\emph{File\_Path, [T$_s$, T$_e$)}) and the valid times indicate
the periods when a file remained unchanged. The Meteo Swiss
dataset\footnote{\scriptsize{Federal Office of Meteorology and
Climatology: \url{http://www.meteoswiss.ch} (2016)}} includes
temperature predictions that have been extracted from the website
of the Swiss Federal Office of Meterology and Climatology. Each
tuple has schema (\emph{Station\_ID, Value\_ID, Value, [T$_s$,
T$_e$)}). The measurements were taken at 80 different
meteorological stations (Station\_ID) in Switzerland from 2005 to
2015 and involve four different metrics (Value\_ID), including
temperature and precipitation. Measurements are 10 minutes apart
and -- in order to produce intervals -- we merged time points
whose measurements differ by less than 0.1.

\begin{table}[!h]
\caption{Real-World Dataset Properties}
\label{tab:realDatasetSummary}
\centering
\fontsize{8pt}{11pt}\selectfont
\setlength{\tabcolsep}{2pt}
\scalebox{1}{
\begin{tabular}{| M{5cm} !{\VRule[1.5pt]} M{1.2cm} |
				  M{1.2cm}  | @{}m{0cm}@{}}
\hline
\textbf{}  &  \textbf{Meteo} &  \textbf{Webkit} & \\  \specialrule{1.2pt}{0pt}{0pt}
Cardinality     & 10.2M  & 1.5M  &  \\ \hline 
Time Range      & 347M   & 7M    &   \\ \hline 
Min. Duration   & 600    & 0.02  &   \\ \hline 
Max. Duration   & 19.3M  & 6M    & \\ \hline 
Avg. Duration   & 152M   & 1.7M  &   \\ \hline 
Num. of Facts   & 80     & 484K  & \\ \hline 
Distinct Points & 545K   & 144K  &   \\ \hline 
Max Num. of Tuples (per time point)	& 140 & 369K &  \\ \hline 
Avg Num. of Tuples (per time point) & 37  & 21   &  \\ \hline 
\end{tabular}}
\end{table}

The main properties of these datasets are summarized in
Table~\ref{tab:realDatasetSummary}. For both datasets we produced
a second relation by shifting the intervals of the original
dataset, without modifying the lengths of the intervals. The
start/end points of the new relation were chosen according 
to the distribution of the original ones. 

\subsection{Runtime}

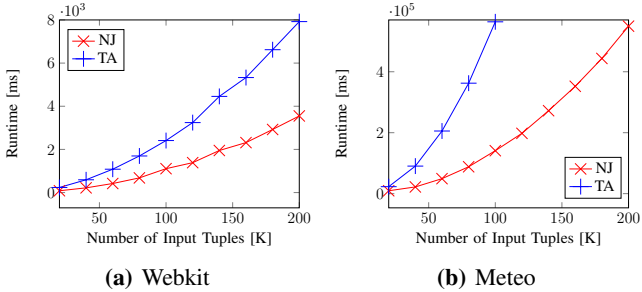
\begin{figure}[!htbp]\centering
\begin{subfigure}[b]{0.48\linewidth}\centering
\scalebox{0.5}{
\begin{tikzpicture}
\begin{axis}[scale only axis,
			 scaled y ticks=base 10:-3,
             height=5cm, 
             width=1.5\linewidth,
             xlabel=Number of Input Tuples {[K]},
             ylabel=Runtime {[ms]},
             label style={font=\large},
             legend style={at={(0.27, 0.95)}, font=\large},
             ymax=8*10^3,
             tick label style={font=\large},
             mark size=4pt,
             xmin = 20,
             xmax = 200]
\addplot[color=red,mark=x,mark size=6pt]
  file [skip first]{figures/webkit_small_wuo_lta.tsv};
\addplot[color=blue,mark=+,mark size=6pt]
  file [skip first] {figures/webkit_small_wuo_norm.tsv};
\legend{NJ,TA}
\end{axis}
\end{tikzpicture} }
\caption{Webkit}
\label{fig:webkitWuoSmall}
\end{subfigure}
\begin{subfigure}[b]{0.48\linewidth}\centering
\scalebox{0.5}{
\begin{tikzpicture}
\begin{axis}[scale only axis,
             height=5cm, 
             width=1.5\linewidth,
             xlabel=Number of Input Tuples {[K]},
             ylabel=Runtime {[ms]},
             label style={font=\large},
             legend style={at={(0.97, 0.27)}, font=\large},
			 ymax = 5.7*10^5,
             tick label style={font=\large},
             mark size=4pt,
             xmin = 20,
             xmax = 200]
\addplot[color=red,mark=x,mark size=6pt]
  file [skip first]{figures/meteo_small_wuo_lta.tsv};
\addplot[color=blue,mark=+,mark size=6pt]
  file [skip first] {figures/meteo_small_wuo_norm.tsv};
\legend{NJ,TA}
\end{axis}
\end{tikzpicture}
}
\caption{Meteo}
\label{fig:meteoWuoSmall}
\end{subfigure}
\caption{W$_{UO}$: Overlapping and Unmatched Windows}
\label{fig:overlapping}
\end{figure}

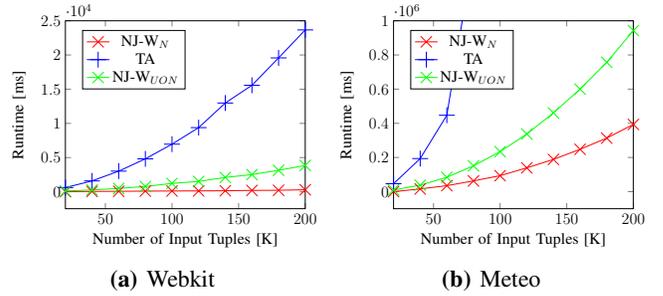
\begin{figure}[!htbp]\centering
\begin{subfigure}[b]{0.48\linewidth}\centering
\scalebox{0.5}{
\begin{tikzpicture}
\begin{axis}[scale only axis,
             height=5cm, 
             width=1.5\linewidth,
             xlabel=Number of Input Tuples {[K]},
             ylabel=Runtime {[ms]},
             label style={font=\large},
             legend style={at={(0.5, 0.95)}, font=\large},
             ymax=2.5*10^4,
             tick label style={font=\large},
             mark size=4pt,
             xmin = 20,
             xmax = 200]
\addplot[color=red,mark=x,mark size=6pt]
  file [skip first]{figures/webkit_small_wn_lta.tsv};
\addplot[color=blue,mark=+,mark size=6pt]
  file [skip first] {figures/webkit_small_wn_norm.tsv};
 \addplot[color=green, mark=x,mark size=6pt]
        file [skip first]{figures/webkit_small_lta.tsv};
\legend{NJ-W$_{N}$,TA, NJ-W$_{UON}$}
\end{axis}
\end{tikzpicture}}
\caption{Webkit}
\label{fig:webkitWnSmall}
\end{subfigure}
\begin{subfigure}[b]{0.48\linewidth}\centering
\scalebox{0.5}{
\begin{tikzpicture}
\begin{axis}[scale only axis,
     		 height=5cm, 
             width=1.5\linewidth,
             xlabel=Number of Input Tuples {[K]},
             ylabel=Runtime {[ms]},
             label style={font=\large},
             legend style={at={(0.5, 0.95)}, font=\large},
			 ymax = 10^6,
             tick label style={font=\large},
             mark size=4pt,
             xmin = 20,
             xmax = 200]
\addplot[color=red,mark=x,mark size=6pt]
  file [skip first]{figures/meteo_small_wn_lta.tsv};
\addplot[color=blue,mark=+,mark size=6pt]
  file [skip first] {figures/meteo_small_wn_norm.tsv};
\addplot[color=green, mark=x,mark size=6pt]
  file [skip first]{figures/meteo_small_lta.tsv};
\legend{NJ-W$_{N}$,TA, NJ-W$_{UON}$}
\end{axis}
\end{tikzpicture}
}
\caption{Meteo}
\label{fig:meteoWnSmall}
\end{subfigure}
\caption{Negating Windows}
\label{fig:negating}
\end{figure}

\begin{figure}[!htbp]\centering
\begin{subfigure}[b]{0.48\linewidth}\centering
\scalebox{0.5}{
\begin{tikzpicture}
\begin{axis}[scale only axis,
             height=5cm,
             width=1.5\linewidth,
             xlabel=Number of Input Tuples {[K]},
             ylabel=Runtime {[ms]},
             label style={font=\large},
             legend style={font=\large},
			 ymax = 10^7,
             tick label style={font=\large},
             mark size=4pt,
             xmin = 20,
             xmax = 200]
\addplot[color=red,mark=x,mark size=6pt]
  file [skip first]{figures/webkit_small_lta.tsv};
\addplot[color=blue,mark=+,mark size=6pt]
  file [skip first] {figures/webkit_small_norm.tsv};
\legend{NJ,TA}
\end{axis}
\end{tikzpicture}}
\caption{Webkit}
\label{fig:webkitLeftOuterJoinSmall}
\end{subfigure}
\begin{subfigure}[b]{0.48\linewidth}
\centering
\scalebox{0.5}{
\begin{tikzpicture}
\begin{axis}[scale only axis,
             height=5cm, 
             width=1.5\linewidth,
             xlabel=Number of Input Tuples {[K]},
             ylabel=Runtime {[ms]},
             label style={font=\large},
             legend style={at={(0.97, 0.27)}, font=\large},
             ymax=10^6,
             tick label style={font=\large},
             mark size=4pt,
             xmin = 20,
             xmax = 200]
\addplot[color=red,mark=x,mark size=6pt]
  file [skip first]{figures/meteo_small_lta.tsv};
\addplot[color=blue,mark=+,mark size=6pt]
  file [skip first] {figures/meteo_small_norm.tsv};
\legend{NJ,TA}
\end{axis}
\end{tikzpicture}
}
\caption{Meteo}
\label{fig:meteoLeftOuterJoinSmall}
\end{subfigure}
\caption{TP Left Outer-Join}
\label{fig:leftjoin}
\end{figure}
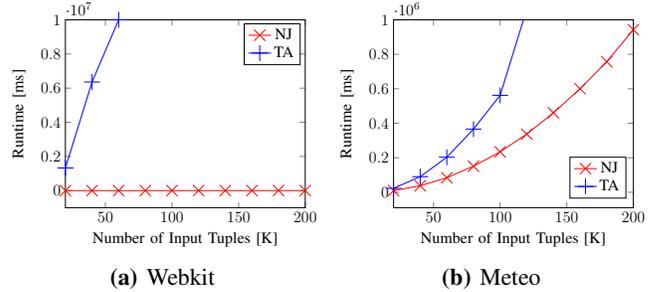

In Fig.~\ref{fig:overlapping}, \ref{fig:negating}, 
\ref{fig:leftjoin} we illustrate the runtime for the overlapping
and unmatched windows, negating windows, and for a TP left
outer join, respectively, over subsets of the Webkit and Meteo
dataset. The subsets range from 20K to 200K tuples. For Webkit
dataset, as a $\theta$ condition we apply equality of the
\emph{File\_Path}, i.e., we combine tuples referring to the same
file. For Meteo dataset, we apply equality on \emph{Value\_IDs}
and inequality on \emph{Station\_IDs}, i.e. we combine tuples
with measurements on the same metric but taken in different 
stations. 

Fig.~\ref{fig:overlapping} shows the runtime of NJ and TA
for the set $\mathtt{w}_\kat{UO}$
(Algorithm~\ref{algo:negatingJoins}), including the unmatched and
overlapping windows. Both approaches follow a similar trend and
the reason is that the most computationally demanding part of
both is a conventional left join, used to identify the pairs of
tuples that overlap. As shown in Fig.~\ref{fig:qTrees}, NJ only
executes this join once whereas TA executes it twice. As a result,
NJ is two to four times faster. 

In Fig.~\ref{fig:negating}, we have illustrated the runtime
for the computation of negating windows. In NJ, negating
windows are computed by applying $LAWA_\kat{N}$ on the set
$\mathtt{w}_\kat{UO}$. Thus, we have illustrated their
computation time both including ($W_{UON}$) and excluding
($W_{N}$) the runtime for $\mathtt{w}_\kat{UO}$. In the case
of $W_{UON}$, NJ computes the negating windows four to
ten times faster than TA whereas, in the case of $W_{N}$,
it computes them twelve to twenty times faster.

Finally, the runtimes of both NJ and TA for a TP left-outer join
are illustrated in Fig.~\ref{fig:leftjoin}. To compute the join
with TA, a duplicate-eliminating is applied on the query trees in
Fig.~\ref{fig:qTreeTAuo} and Fig.~\ref{fig:qTreeTAn} to combined
the partial results and remove the redundant unmatched windows. 
Its runtime for the TP left-outer join is much higher than the sum
of the runtimes of the windows as presented in
Fig.~\ref{fig:overlapping} and Fig.~\ref{fig:negating}. The
reason for that is that when the union of the query trees in 
Fig.~\ref{fig:qTreeTAuo} and \ref{fig:qTreeTAn} is performed,
the $\theta$ condition of the TP join is ignored for the right 
subtree of Fig.~\ref{fig:negating}. The optimizer opts for a
nested loop for its computation and this takes a huge toll on
TA's runtime making NJ two orders of magnitude faster. 

Meteo dataset contains a number of distinct values much smaller
than its size, an analogy maintained in the subsets due to the
use of the uniform distribution in their creation. As a result,
the condition is not very selective and the runtime of both NJ
and TA is higher than it was in the case of the webkit dataset.
In all cases, the runtime of NJ outperforms TA by four to ten
times.

\subsection{Runtime Breakdown  and Scalability}

The query tree of the NJ approach (cf. Fig.~\ref{fig:qTreeLAWA}) 
consists of the nodes $\LJoin_{\theta \land \theta_{o}}$, 
$\mathcal{W}_{uo}$ and $\mathcal{W}_{n}$ nodes. The way that
the node $\LJoin_{\theta \land \theta_{o}}$ is computed is
completely determined by PostgreSQL's optimizer, given the
condition applied on the non-temporal attributes. The most 
demanding part of the node $\mathcal{W}_{n}$ is handling
the tuples valid over the interval of the window. In
Fig.~\ref{fig:runtimeBreakdown}, we breakdown the runtime of a
TP left outer join on the percentage occupied by each node of
the query tree for Webkit and Meteo dataset, respectively.
As shown in the graphs, the conventional left-outer join (CLJ)
occupies most of the runtime of the TP left outer join (NJ) which
is more than 50\% for Webkit dataset. The calls to LAWA$_U$ and
LAWA$_N$, for the computation of the nodes $\mathcal{W}_{uo}$ and
$\mathcal{W}_{n}$ respectively, correspond to a small percentage
of the runtime in Webkit dataset. However, they tend to be more
time-consuming for Meteo dataset. This behaviour lies in the
dataset characteristics and in the query performed. In meteo, 
the $\theta$ condition used requests for the tuples combined to
have the same metric but to refer to different stations. 
Measurements over all stations take place at similar times
and, for multiple output intervals, all valid tuples might
contribute in the output, making the computations much more
demanding. 

\input{8c_stage_analysis_and_scalability}

NJ is the only scalable approach integrated in PostgreSQL that
can be used for the computation of all TP joins including
negation. In Fig.~\ref{fig:Scalability}, we depict the
performance of NJ for the computation of a TP left outer join
for larger subsets of the webkit and meteo datasets. \emph{TA} is
not taken into consideration, since its runtimes were already one
to four orders of magnitude higher than NJ's when applied on the
smaller datasets. The dataset sizes vary from 100K to 1M tuples.
NJ's implementation is based on a conventional left outer join 
and its performance is influenced by the condition on the
non-temporal attributes, since the optimizer opts for a different
type of join. The selectivity of the condition applied in the
webkit dataset is higher, allowing for the computation of the
left outer join using a merge join. On the contrary, in the case
of meteo dataset, a nested loop has to be computed. As a result,
NJ scales more efficiently when applied on the webkit dataset,
with its runtime being two minutes on average and always less
than five minutes for datasets less than 2M. 

\begin{figure}[!htbp]
\centering
\scalebox{0.5}{
\begin{tikzpicture}
\begin{axis}[scale only axis,
             height=4cm,
             width=1.5\linewidth,
             xlabel=Number of Input Tuples {[M]},
             ylabel=Runtime {[ms]},
             label style={font=\large},
             legend style={font=\large},
			 ymax=10^6,
             tick label style={font=\large},
             mark size=4pt,
             xmin = 0.1,
             xmax = 2]
\addplot[color=red,mark=x,mark size=6pt]
         file [skip first]{figures/webkit_large_lta.tsv};
\addplot[color=blue,mark=+,mark size=6pt]
         file [skip first] {figures/meteo_large_lta.tsv};
\legend{Webkit,Meteo}
\end{axis}
\end{tikzpicture}
}
\caption{Scalability}
\label{fig:Scalability}
\end{figure}

%% file: 8c_stage_analysis_and_scalability.tex
\begin{figure}[!htbp]
\center
\begin{subfigure}[b]{0.48\linewidth}
\centering
\scalebox{0.7}{
\begin{tikzpicture}
\begin{axis}[legend columns=-1, 
			 ylabel near ticks,
			 xlabel near ticks,
			 label style={font=\small},
			 tick label style={font=\small} ,
			 legend style={nodes={scale=0.7, transform shape},
			 			   at={(1.05, 1.27)} },
			 xlabel=Number of Input Tuples {[K]}, 
		 	 ylabel=Runtime Percentage {[ms]},
			 bar width = 4pt,
	  		 scale only axis,
			 height=3.5cm,
			 width=\linewidth,
			 xtick={40,80,120,160,200},
			 ybar=2.2*\pgflinewidth,
			 ymax=120,
			 enlargelimits=0.1]	
\addplot +[area legend] coordinates {(40,100) (80,100) (120,100) (160,100) (200,100)};
\addplot +[area legend] coordinates {(40,65.828) (80,80.110) (120,83.118) (160,86.519) (200,87.714)};
\addplot +[area legend] coordinates {(40,14.877) (80,7.411) (120,8.394) (160,5.451) (200,4.514)};
\addplot +[area legend] coordinates {(40,19.295) (80,12.479) (120,8.488) (160,8.030) (200,7.772)};
\legend{NJ, CLJ, WUO, WN}  
\end{axis}
\end{tikzpicture}}
\caption{Webkit Dataset.}
\label{fig:stageWebkit}
\end{subfigure}
\begin{subfigure}[b]{0.48\linewidth}
\centering
\scalebox{0.7}{
\begin{tikzpicture}
\begin{axis}[legend columns=-1, 
			 ylabel near ticks,
			 xlabel near ticks,
			 legend style={nodes={scale=0.7, transform shape},
			 			   at={(1.05, 1.27)} },
			 xlabel=Number of Input Tuples {[K]}, 
			 ylabel=Runtime Percentage {[ms]},
			 bar width = 4pt,
			 scale only axis,
			 height=3.5cm,
			 width=\linewidth,
			 xtick={40,80,120,160,200},
			 ybar=2.2*\pgflinewidth,
			 ymax=120,
			 enlargelimits=0.1]			 
\addplot +[area legend] coordinates {(40,100) (80,100) (120,100) (160,100) (200,100)};
\addplot +[area legend] coordinates {(40,57.627) (80,58.138) (120,57.835) (160,57.859) (200,57.870)};
\addplot +[area legend]  coordinates {(40,42.732) (80,41.809) (120,0.097) (160,0.980) (200,0.528)};
\addplot +[area legend]  coordinates {(40,0) (80,0) (120,42.068) (160,41.165) (200,41.601)};
\legend{NJ, CLJ, WUO, WN}  
\end{axis}
\end{tikzpicture}}
\caption{Meteo Dataset.}
\label{fig:stageMeteo}
\end{subfigure}

\caption{Runtime Breakdown. CLJ is $\LJoin_{\theta \land 
\theta_{o}}$ and NJ is $\LJoin_{\theta}^{Tp}$ .}
\label{fig:runtimeBreakdown}
\end{figure}